\documentclass[pra,aps,a4paper,twocolumn,repreprint,superscriptaddress,longbibliography]{revtex4-2}

\usepackage{amssymb}
\usepackage{amsmath}
\usepackage{amsfonts}
\usepackage{natbib}
\usepackage{hyperref}
\usepackage{graphicx}
\usepackage[dvipsnames]{xcolor}
\usepackage{bm}

\DeclareMathOperator{\re}{Re}

\DeclareMathOperator{\sign}{sign}
\DeclareMathOperator{\popcorn}{popcorn}

\def\br{\mathbf{r}}

\def\vp{\varphi}
\def\kp{k}
\def\ve{\varepsilon}

\newcommand{\eps}{\varepsilon}
\newcommand{\corr}[1]{\langle #1\rangle}

\def\be{\begin{equation}}
\def\ee{\end{equation}}

\begin{document}

\title{
Vortex core near planar defects in a clean layered superconductor
}

\author{Uliana E. Khodaeva}
\affiliation{Skolkovo Institute of Science and Technology, Moscow 121205, Russia}
\affiliation{Moscow Institute of Physics and Technology, Moscow 141700, Russia}
\affiliation{L. D. Landau Institute for Theoretical Physics, Chernogolovka 142432, Russia}

\author{Mikhail A. Skvortsov}
\affiliation{L. D. Landau Institute for Theoretical Physics, Chernogolovka 142432, Russia}

\begin{abstract}
We investigate the structure of quasiparticle states localized in a core of an Abrikosov vortex in a clean layered superconductor in the presence of planar defects. It is shown that even a highly transparent defect opens a minigap at the Fermi energy. Its magnitude, $E_g\sim\Delta\sqrt{R}$, exceeds the mean level spacing for the chiral branch, $\omega_0\sim\Delta^2/E_F$, already for very small values of the reflection coefficient off the defect, $R\ll1$ ($\Delta$ is the bulk gap). For $R\gtrsim\sqrt{\Delta/E_F}$, formation of the minigap is accompanied by the appearance of subgap states localized along the defect, in accordance with [A. V. Samokhvalov \emph{et al.}, Phys.\ Rev.\ B \textbf{102}, 174501 (2020)]. The minigap takes its maximal value for the vortex located right at the defect, decreases with increasing the distance $b$ from the defect, and closes when $k_Fb\sim (\Delta/\omega_0)\sqrt{R}$. We also study various configurations of several planar defects (few crossing planes, stars, periodic structures). Although the minigap remains, a strong commensurability effect is observed. For two crossing planar defects, the magnitude of the minigap strongly depends on how close the intersection angle is to a rational fraction of $\pi$.

\end{abstract}

\date{\today} 

\maketitle

\section{Introduction}

Vortices are responsible for low-frequency dissipation in the mixed state of $s$-wave superconductors at low temperatures \cite{BS}. When the temperature $T$ drops well below the superconducting gap $\Delta$ and bulk quasiparticle excitations are frozen out, entropy transfer may take place only in the vortex core. The latter can be considered as a piece of a normal metal with the size of the coherence length $\xi$ and finite density of states (DOS) at the Fermi level. Microscopic justification of this picture was provided by Caroli, de Gennes, and Matricon (CdGM) \cite{CdGM}, who calculated the spectrum of excitations localized in the vortex core. They obtained that in the absence of impurities the low-energy spectral branch is given by a dense set of equidistant levels,
\be
\label{CdGM-spectrum}
  E_{\mu} = \mu\omega_0 ,
\ee
with the level spacing $\omega_0\sim\Delta^2/E_F\ll\Delta$ and half-integer $\mu$ ($E_F$ is the Fermi energy and we set $\hbar=1$). Equation \eqref{CdGM-spectrum} holds in the two-dimensional (2D) case applicable for pancake vortices in layered superconductors. Account of motion along the vortex (in $z$ direction) broadens each level $E_\mu$ into a zone dependent on the momentum $k_z$. For strongly anisotropic superconductors such a broadening is small and can be neglected.

The parameter $\mu$ in Eq.\ \eqref{CdGM-spectrum} is a half-sum of the angular momenta of the electron and hole components of the excitation. If one relaxes the constraint $\mu=\mathbb{Z}+1/2$ and treats $\mu$ as a continuous variable then the chiral branch can be considered as a gapless fermionic zero mode \cite{Volovik-book}. Its existence is protected by topological arguments \cite{Volovik1993} and physically is related to vanishing of the average order parameter seen by trapped particles due to $2\pi$ winding of its phase \cite{AZ}.

Disorder typically modeled by point-like impurities breaks the axial symmetry of the problem and leads to mixing of the chiral states. That calls for a statistical description, which has been extensively studied in  the clean ($1/\tau\ll\omega_0$) \cite{LO98}, moderately clean ($\omega_0\ll1/\tau\ll\Delta$) \cite{SKF98,Koulakov,BHL99,Fujita}, and dirty ($\Delta\ll1/\tau$) \cite{Bundschuh99} limits, where $\tau$ is the elastic scattering time. Despite several types of the spectral statistics have been identified, the coarse-grained DOS averaged over a window larger than $\omega_0$ is still $1/\omega_0$, indicating that the chiral branch in the presence of point-like impurities remains (quasiclassically) gapless.

When a vortex starts moving driven by an electric current, impurity potential in its core is being changed with time. This is the origin of the spectral flow along the chiral branch, leading to the heating of the vortex core and eventually to the energy dissipation. On a quasiclassical level, the theory of flux-flow conductivity has been developed in Refs.\ \cite{GorkovKopnin1973,KopninKravtsov1976,LO-review}. Peculiar effects related to spectrum discreteness were studied in Refs.\ \cite{LO98,Koulakov,Guinea,FS97,SF00,SIB03}.

A different type of imperfections is provided by extended defects, such as columnar defects
\cite{Melnikov09,Shapiro,Melnikov15}. Those structures have a much more pronounced effect on the chiral branch as they completely take low-$\mu$ excitations out of the game, leading to the formation of the minigap $E_g\sim\omega_0k_Fb$, where $b$ is the radius of the columnar defect and $k_F$ is the Fermi momentum. Other types of extended defects relevant for vortex pinning and microwave absorption are grain boundaries \cite{bound1,bound2}, twin boundaries \cite{twin1,twin2}, anti-phase boundaries \cite{antiphase}, etc.

In a recent paper, Samokhvalov \emph{et al.}\ \cite{Melnikov2020} considered modification of the vortex-core states in a clean superconductor in the presence of a sufficiently weak planar defect with the normal-incidence reflection coefficient $R\ll1$ passing through the vortex center. Solving the Bogolyubov--de Gennes equation in the quasiclassical approximation, they obtained that the defect breaks the continuity of the chiral branch and opens a minigap $E_g\gg\omega_0$ in its spectrum, which grows with the strength of the defect, $R$. Another prediction of Ref.\ \cite{Melnikov2020} is the existence of the states localized along the defect, which appear through a topological transition with increasing $R$ above $1/k_F\xi$. The resulting DOS structure in this regime is quite complicated. It is characterized by the minigap $E_g\sim\Delta\sqrt{R}$ for the majority of states and the presence of a number of sub-minigap states referred to as a ``soft gap'' in Ref.\ 
\cite{Melnikov2020}.

Although quasiclassical approximation is a standard tool in vortex physics \cite{Kopnin-book}, it should be applied with care. A known issue is a controversy on the presence of a hard gap in a normal diffusive metal proximitized by a superconductor: While the microscopic approach based on the Usadel equation predicts a hard gap of the order of the Thouless energy \cite{Golubov89,Zhou98,Ostrovsky01}, a trajectory-based approach leads to a soft gap \cite{Melnikov_SNS}. The origin of the discrepancy can be traced back to the absence of quantum transitions between trajectories in the quasiclassical treatment. Such transitions automatically incorporated in the Usadel equation 
become essential in the low-energy (long-time) limit and eventually lead to the hard gap formation \cite{Taras01}.

In this paper we consider the effect of weak planar
defects on the quasiparticle excitations localized in the vortex core in a clean superconductor (no impurities). Instead of relying on the quasiclassical approach, we develop a fully quantum-mechanical approach based on the one-dimensional (1D) nature of the chiral branch. Then the knowledge of exact clean wave functions \cite{FS97} allows us to calculate the matrix elements of the defect without assuming it to be small. The resulting 1D quantum mechanics is solved either in the momentum representation or in the dual angular representation.

In order to simplify the analysis, we consider the case of layered superconductors. Then the problem of a vortex near a planar defect reduces to the 2D problem of a \emph{pancake} vortex near a \emph{linear} defect. Assuming such a 2D geometry, hereinafter planar defects are referred to as linear ones.

\begin{figure}
\centerline{\includegraphics[width=1.0\columnwidth]{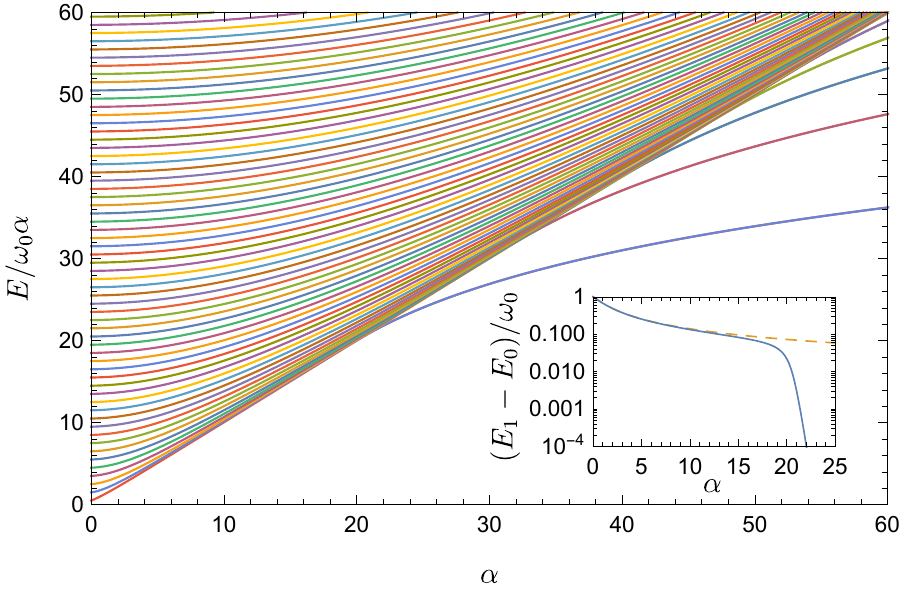}}
\caption{Positive energy levels $E_n$ for a vortex with $k_F\xi=200$ as a function of the dimensionless defect strength $\alpha$. One can clearly see a linearly growing gap $E_g=\omega_0\alpha$.
At $\alpha \gtrsim 20$ some pairs of levels become nearly degenerate and detach from the gapped majority of states. 
Inset shows the energy difference between the lowest pair of levels, with the sharp drop indicating the emergence of the first pair of nearly degenerate subgap states (indistinguishable at the main panel) at $\alpha \simeq 20$.}
\label{F:levels-M}
\end{figure}

For a single linear defect passing through the vortex center, we reproduce and clarify the results obtained by the semiclassical trajectory approach \cite{Melnikov2020}. Characterizing the defect strength by a dimensionless parameter $\alpha\sim(k_F\xi)\sqrt{R}$ [for a precise definition, see Eq.\ \eqref{alpha_kappa}], we identify two regimes:
\begin{itemize}
\item Weak defects with $\alpha\ll\sqrt{k_F\xi}$ open a minigap $E_g\approx\alpha\omega_0$. In this case the whole spectrum can be determined analytically.
\item For stronger defects with $\alpha\gg\sqrt{k_F\xi}$, the main part of the spectrum is gapped with $E_g\approx\alpha\omega_0$, but a number of subgap states appear at energies $E_n<E_g$. Those are the ``soft-gap'' states of Ref.\ \cite{Melnikov2020}.
\end{itemize}
This picture is illustrated in Fig.\ \ref{F:levels-M}, where we plot the spectrum of localized states as a function of $\alpha$ obtained numerically for a vortex with $k_F\xi=200$. At $\alpha=0$ we have an equidistant set of CdGM levels \eqref{CdGM-spectrum}.
One can clearly see the opening of the gap $E_g\approx\alpha\omega_0$ accompanied by sequential splitting of states with a weaker $\alpha$-dependence starting at $\alpha\approx20\approx\sqrt{2k_F\xi}$. Since the majority of states are gapped with $E_g\approx\alpha\omega_0$, we would like to refer to those new states as subgap states refraining from using the ``soft-gap'' terminology. 

Inset to Fig.\ \ref{F:levels-M} shows the difference $E_1-E_0$ between the first two levels. First it decreases nearly as $E_1-E_0\sim\omega_0/\sqrt\alpha$ due to level crowding above the minigap, with the dashed line being the exact expression obtained from Eqs.\ \eqref{eq-spectrum-E} and \eqref{eq-spectrum}. However at $\alpha\approx20$ two lowest levels nearly merge, with the splitting decaying exponentially. This corresponds to an exponentially weak hybridization of a pair of states localized on both sides of the defect.

We also analyze a number of more complicated configurations of linear defects in the vortex core, assuming $\alpha\lesssim\sqrt{k_F\xi}$, where a simple quantum-mechanical description can be developed.
In order to study the robustness of the minigap, we consider a vortex displaced from the linear defect at a finite distance $b$. We obtain that the minigap $E_g(\alpha, b) \approx (\alpha-k_Fb)\omega_0$ exists as long as $k_Fb<\alpha$.
For several intersecting linear defects, we report on a pronounced commensurability effect, with the magnitude of the minigap being highly sensitive to the angle between the defects and the fraction it constitutes with $\pi$. The maximal value the minigap can take is again $E_g\sim\alpha\omega_0$, whereas for incommensurate angles it still survives and is bounded from below by $E_g\sim\sqrt\alpha\omega_0$.
Finally, we demonstrate that the minigap persists also for periodic structure of linear defects, though its value decreases with decreasing the period of the defect lattice.

The paper is organized as follows. In Sec.\ \ref{S:one-line} we map the problem of the excitation spectrum in the core of a 2D pancake vortex placed near a linear defect to a simple 1D quantum mechanics, demonstrate gap opening, and determine its dependence on the defect strength and the distance between the vortex center and the defect.
Peculiar incommensurability effects arising at intersecting multiple defects configurations are analyzed in Sec.\ \ref{S:multiple}. Periodic structures of defects are studied in Sec.\ \ref{S:Periodic}. While all the above results were obtained for $\alpha\lesssim\sqrt{k_F\xi}$, in Sec.\ \ref{S:NN} we discuss the appearance of the subgap states at $\alpha\gtrsim\sqrt{k_F\xi}$ and relation to Ref.\ \cite{Melnikov2020}. In Sec.\ \ref{S:self_consistency} we consider the effect of the self-consistent determination of the order parameter and show that it can be neglected. The results obtained are discussed in Sec.\ \ref{S:Conclusion}. Important technical details are relegated to several Appendixes.

\section{Pancake vortex near a linear defect}
\label{S:one-line}

In this section we consider the case of a linear defect, passing at a distance of $b$ from the vortex center. The defect is assumed to be weak enough, such that it does not break the superconductor into two weakly coupled pieces and cannot be described by the tunneling Hamiltonian approximation. Instead, we model it by a delta-function potential \cite{BTK,Melnikov2020}
\begin{equation}
\label{V-ini}
  V(\br) = \frac{\hbar^2\varkappa}{m}
\delta(r_1 - b) ,
\end{equation}
where $\br=(r_1,r_2)$ and $m$ is the electron's mass (we assume parabolic dispersion). The strength of the defect specified by the parameter $\varkappa$ can be conveniently characterized by the normal-incidence reflection coefficient:
\be
\label{R-def}
R = \frac{\varkappa^2}{k_F^2 + \varkappa^2}.
\ee
The defect is weak provided $\varkappa \ll k_F$ and hence $R\ll1$. Below the influence of the defect on the chiral states will be described by the dimensionless parameter $\alpha$ introduced in Eq.\ \eqref{alpha_kappa}.

\subsection{CdGM states}

Here, we summarize relevant information about quasiparticle states localized in the vortex core in a clean 2D superconductor \cite{CdGM}. They are obtained as eigenstates of the Bogoliubov--de Gennes (BdG) equation \cite{deGen}
\begin{equation}\label{BdG}
  \mathcal{H}(\br)\Psi(\br) = \varepsilon\Psi(\br) 
\end{equation}
for a two-component (particle/hole) wave function 
$\Psi(\br)$. Choosing the vortex order parameter in the form $\Delta(\br)=|\Delta(r)|e^{-i\vp}$, where $r$ and $\vp$ are polar coordinates, the BdG Hamiltonian can be written as
\be
\label{H-BdG}
  \mathcal{H}(\br) = \begin{pmatrix}
    H & \Delta(\br)\\
    \Delta^{*}(\br) & -H^{*}\\
  \end{pmatrix},
\ee
where $\Delta(\br)$ is the superconducting order parameter and $H$ is the single-particle Hamiltonian:
\be
H = \frac{1}{2m}\left(\mathbf{p} - \frac{e}{c}\mathbf{A} \right)^2 - E_F.
\ee
Here $E_F$ is the Fermi energy and $\mathbf{A}$ is the  vector potential (that will be neglected assuming strong type-II superconductivity).

The low-lying spectral branch of this equation given by Eq.~\eqref{CdGM-spectrum} was obtained by CdGM \cite{CdGM}, who worked in a quasiclassical approximation. The wave function of the $\mu$th state valid for all $\mu$ has the form \cite{SKF98}
\begin{equation}\label{psi_mu}
  \Psi_{\mu}(\mathbf{r}) 
  = 
  Ae^{-K(r)}
  \begin{pmatrix}
  J_{\mu - 1/2}(k_{F}r) \, e^{i(\mu-1/2)\vp} \\
  J_{\mu + 1/2}(k_{F}r) \, e^{i(\mu+1/2)\vp} \\
  \end{pmatrix},
\end{equation}
where the envelope $e^{-K(r)}$ with
\begin{equation}
\label{K-def}
K(r) = \frac{1}{\hbar v_{F}}\int_{0}^{r}\Delta(r')\,dr'
\end{equation}
decays exponentially at $r\gg\xi$,
and $A$ is the normalization factor:
\begin{equation}
A^{2} = \left[\frac{4}{k_{F}}\int_{0}^{\infty}e^{-2K(r)}\,dr \right]^{-1} \sim \frac{k_{F}}{\xi}.
\end{equation}
Strictly speaking, the factor $A$ also depends on $\mu$, but this dependence can be neglected for $E_\mu\ll\Delta$.

The level spacing $\omega_0\sim\Delta^2/E_F\ll\Delta$ is determined by the profile of the order parameter $\Delta(r)$:
\begin{equation}
\label{omega0}
  \omega_{0} 
  = 
  \frac{\int_{0}^{\infty}\frac{\Delta(r)}{k_{F}r}e^{-2K(r)}\,dr}
  {\int_{0}^{\infty}e^{-2K(r)}\,dr} .
\end{equation}

\subsection{Projection to the chiral branch}
\label{SS:matelems}

Recognition that all low-energy states in the vortex core are exhausted by the CdGM chiral branch is vital for describing of quasiparticle rearrangement by a weak potential perturbation. For a finite $V(\br)$, it makes it possible to reduce a complicated BdG equation \eqref{BdG}, which is a matrix differential equation in 2D, to a much simpler 1D problem by projecting it onto the states of the chiral branch. In such an approach pioneered in Ref.\ \cite{SKF98}, the BdG Hamiltonian is mapped onto a matrix
\be
\label{Hmunu}
  H_{\mu\nu} 
  = 
  \mu 
  \omega_0
  \delta_{\mu\nu} + V_{\mu\nu} ,
\ee
where $V_{\mu\nu}$ is the matrix element of the potential $V(\br)$ in the chiral basis \eqref{psi_mu}:
\be
\label{V_def}
  V_{\mu\nu} 
  = 
  \int d^2r \,
  \Psi_\mu^+(\br) \tau_3 \Psi_\nu(\br) V(\br),
\ee
Knowledge of the wave functions in an explicit form \eqref{psi_mu} then allows one to calculate $V_{\mu\nu}$, thus obtaining an accurate quantum-mechanical description of the low-energy states in the core, free of any approximations, controlled or uncontrolled.

Reduction to the chiral branch is justified, provided the energy scale of the perturbation $V_{\mu\nu}$ is smaller than the bulk gap (i.e., for disordered superconductors in the clean limit, $1/\tau\ll\Delta$ \cite{SKF98,LO98,SKF98,Koulakov}). Otherwise, mixing of states that do not belong to the chiral branch cannot be avoided, which requires the use of more sophisticated techniques (that is the dirty limit, $\Delta\ll1/\tau$ \cite{LO-review,Bundschuh99}).

The form of the clean eigenfunctions \eqref{psi_mu} suggests \cite{SKF98} switching to a dual representation:
\be
  \psi(x) = \sum_\mu \psi_\mu e^{i\mu x} ,
\ee
with wave functions depending on an angular variable $x$. In physical terms, this angle describes Andreev precession of a quasiclassical trajectory in the vortex core \cite{Kopnin-book}. Since angular momenta are half-integer, wave functions in the angular representation must be $2\pi$ antiperiodic: 
\be
\label{psi-2pi}
  \psi(x+2\pi) = -\psi(x).
\ee

In the dual representation, the Hamiltonian given by
\begin{equation}\label{fourier}
  H(x,y) =
  \sum_{\mu\nu} H_{\mu\nu} e^{ix\mu-iy\nu} 
\end{equation}
acquires the form
\be
\label{H(x,y)}
  H(x,y)
  =
  -i \delta(x-y) 
  \omega_0
  \partial_y + V(x,y) .
\ee
Here the first term is due to the abovementioned Andreev precession, whereas the second term is the integral kernel describing quasiparticle scattering from the potential $V(\br)$. For the linear defect with the potential given by Eq.\ \eqref{V-ini}, the corresponding kernel is calculated in Appendix~\ref{V_x_y}. 

In the limit $\alpha\lesssim\sqrt{k_F\xi}$ 
it has the following form:
\begin{gather}
\label{Vxy_b}
  V(x,y)
  =
  i
  \alpha
  \omega_0
  s(x) 
  \times 2\pi \delta(x+y) ,
\\
  s(x)
  =
  e^{2ik_Fb\sin x} \sign(\sin x) .
\end{gather}
Here we introduced a convenient dimensionless defect strength $\alpha$ defined as
\begin{equation}\label{alpha_kappa}
  \alpha 
  = 
  \frac{2\hbar^2\varkappa A^2}{m k_F\omega_0}
  \sim
  k_F\xi \, \sqrt{R}
  ,
\end{equation}
where $R$ is the normal-incidence reflection coefficient of the defect [see Eq.\ \eqref{R-def}].

In the limit $\alpha\gtrsim\sqrt{k_F\xi}$ the width of the delta function in Eq.\ \eqref{Vxy_b} should be taken into account that leads to the formation of subgap states propagating along the defect, as discussed in Sec.\ \ref{S:NN}.

\subsection{Spectral equation}
\label{shred_eq_text}

In the angular representation, the eigenvalue equation is generally of an integrodifferential form.
However, the fact that the kernel $V(x,y)$ given by Eq.\ \eqref{Vxy_b} contains a delta function $\delta(x+y)$ implies that the corresponding Schr\"odinger equation for the function $\psi(x)$ takes a simple quasi-local form:
\begin{equation}\label{shred_b_psi}
    -i\partial_x \psi(x) + i\alpha s(x) \psi(-x) 
    = (E/\omega_0)\psi(x) ,
\end{equation}
where the effect of nonlocality is the admixture of $\psi(-x)$ to the chiral evolution of $\psi(x)$.

Such a structure of the Schr\"odinger equation suggests that it can be reduced to a truly local form by combining $\psi(x)$ and $\psi(-x)$ into a single 2-vector (spinor)
\be
\label{spinor}
  \Psi(x) 
  =
  \begin{pmatrix}
    \psi(x) \\ \psi(-x)
  \end{pmatrix}
,
\ee 
a procedure resembling the Bogoliubov transformation.
Hence, one can rewrite Eq.\ \eqref{shred_b_psi} as an eigenvalue equation
\be
\label{L-eq}
  \begin{pmatrix}
    -i\partial_x & i\alpha s(x) \\
    i\alpha s(-x) & i\partial_x
  \end{pmatrix}  
  \Psi(x) 
  = 
  (E/\omega_0) \Psi(x)
\ee
for a certain differential matrix operator.

Due to $2\pi$ antiperiodicity of $\psi(x)$ it is sufficient to consider Eq.\ \eqref{L-eq} at the interval $x\in[0,\pi]$. By construction, the spinor $\Psi$ obeys the following constraints at the boundaries of this interval:
\be
\label{psi-bc}
  \Psi(0) 
  = 
  \psi(0) \, |+\rangle
, 
\qquad
  \Psi(\pi) 
  = 
  \psi(\pi) \, |-\rangle .
\ee 
with
\be
\label{+-}
  |+\rangle
  = 
  \begin{pmatrix} 1 \\ 1 \end{pmatrix}
, 
\qquad
  |-\rangle
  = 
  \begin{pmatrix} 1 \\ -1 \end{pmatrix}
  .
\ee 

Isolating the first derivative, one can rewrite Eq.\ \eqref{L-eq} as an evolutionary equation
\be
\label{M-eq}
  \partial_x
  \Psi(x) 
  = 
  M_E(x)
  \Psi(x)
\ee
with the $x$-dependent matrix
\be
\label{M-def}
  M_E(x)
  =
  \begin{pmatrix}
    iE/\omega_0 & \alpha e^{2ik_Fb\sin x} \\
    \alpha e^{-2ik_Fb\sin x} & -iE/\omega_0
  \end{pmatrix}
  .
\ee
The solution of the first-order differential equation \eqref{M-eq} can be written as
\be
\label{Psi(x)}
  \Psi(x) 
  = 
  S_E(x)
  \Psi(0) ,
\ee
where $S_E(x)$ is a time-ordered matrix exponent:
\be
\label{S-T}
  S_E(x)
  =
  \text{T}\exp \left[ \int_0^x M_E(y) dy \right] .
\ee

Now using the boundary conditions \eqref{psi-bc} and utilizing the orthogonality of the spinors \eqref{+-}, we arrive at the spectral equation
\be
\label{spectrum-gen}
  \corr{+|S_E(\pi)|+} = 0 ,
\ee
which determines the eigenvalues $E$ for the quasiparticle states in the core of the vortex located near the planar defect.
Unfortunately, for a finite distance between the vortex and the defect ($b\neq0$) an explicit $x$ dependence of the matrix $M_E(x)$ does not allow the $T$\,exponent in Eq.\ \eqref{S-T} to be calculated analytically.
The latter can be done only in the case $b=0$ (vortex right at the defect), which is considered below. The general situation is discussed in Sec.\ \ref{SS:b>0}.

\begin{figure}
\centerline{\includegraphics[width=0.95\columnwidth]{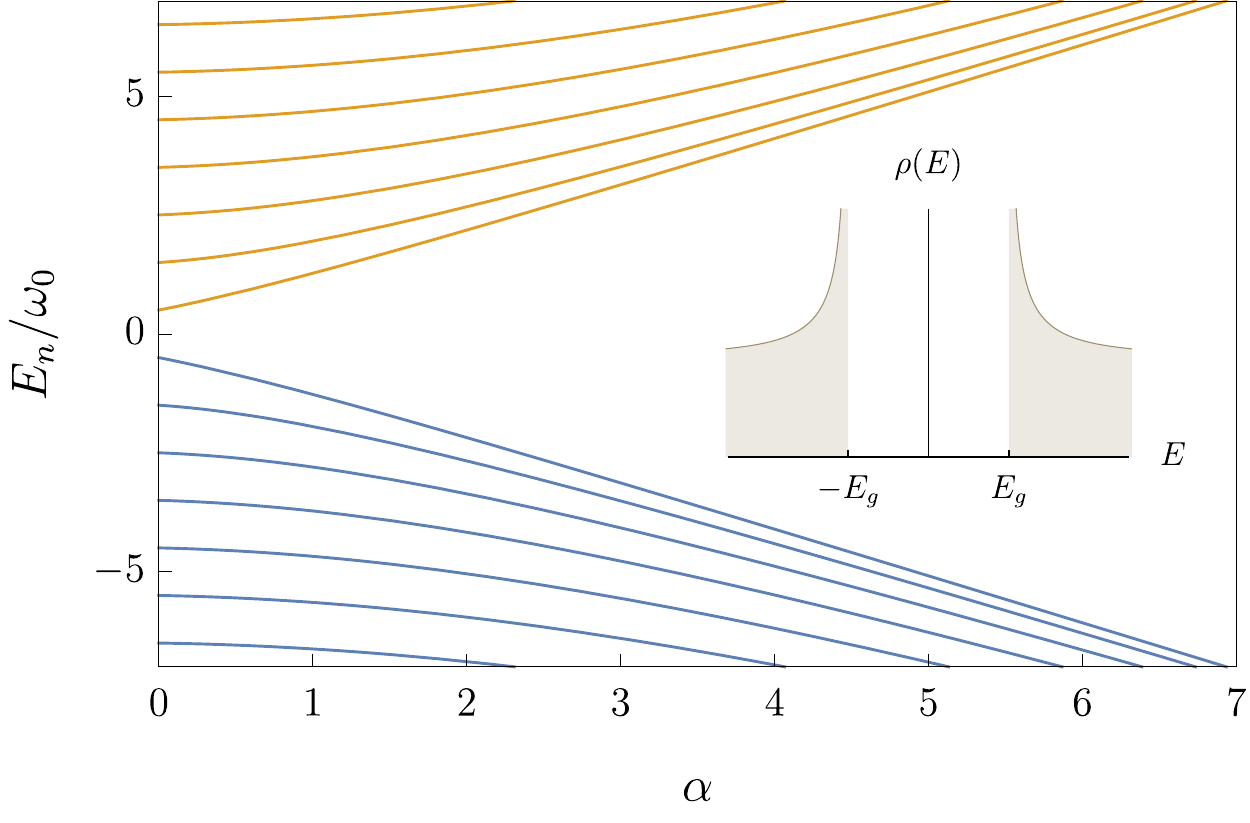}}
\caption{Energy levels for the vortex right at the defect as a function of its strength $\alpha$ obtained from Eqs.\ \eqref{eq-spectrum-E} and \eqref{eq-spectrum}. Opening of the minigap $E_g(\alpha)\approx\alpha\omega_0$ is clearly seen for $\alpha\gg1$. Redistribution of the states of the chiral branch is described by the BCS-type coarse-grained DOS \eqref{DOS} shown in the inset.
}
\label{F:alpha_levels}
\end{figure}

\subsection{Vortex right at the defect ($b=0$)}
\label{zero_distance}

In this section we provide an explicit expression for the spectrum and wave functions in the case of the linear defect passing exactly through the center of the vortex. Although our analysis will be based on the Schr\"odinger equation \eqref{shred_b_psi} in the dual angular representation, it is instructive to give here also the matrix elements of the defect line in the original momentum representation:
\be
\label{V0-munu}
  V_{\mu\nu}^{(0)}
  = 
  \frac{2\alpha\omega_0h_{\mu\nu}}{\pi(\mu+\nu)} ,
\ee
where $h_{\mu\nu} = 1$ for odd $\mu+\nu$ and $h_{\mu\nu} = 0$ for even $\mu+\nu$.
The matrix $V_{\mu\nu}^{(0)}$ can be considered as a generalization of the Hilbert matrix $H_{ij}=1/(i+j-1)$ with $i,j=1,\dots,n$ to two-side infinite case $i,j=-\infty,\dots,\infty$.

\subsubsection{Exact spectrum and gap opening}

At $b=0$ the matrix $M_E(x)$ defined in Eq.\ \eqref{M-def} becomes $x$ independent: $M_E=i(E/\omega_0)\sigma_3+\alpha\sigma_1$, with $\sigma_i$ being the Pauli matrices. The corresponding transfer matrix $S_E(x)$ in Eq.\ \eqref{S-T} is readily calculated and Eq.\ \eqref{spectrum-gen} then provides the energy spectrum. It can be conveniently represented as
\be
\label{eq-spectrum-E}
  E = \omega_0\sqrt{\kp^2 + \alpha^2} \, \sign\kp ,
\ee
where $\kp$ is a real number satisfying the following transcendental equation:
\be
\label{eq-spectrum}
  \alpha + \kp \cot\pi\kp = 0.
\ee
The number $k$ has the physical meaning of momentum, as it becomes clear from the explicit form of the wave function \eqref{wf}.

Equation \eqref{eq-spectrum} defines a discrete set of allowed momenta $k_n(\alpha)$ placed symmetrically around zero. For convenience we consider below only positive $k_n$, which we label starting with $n=0$. Though depending on $\alpha$, each $k_n$ belongs to a small window
\be
\label{kn-window}
  n+1/2 \leq k_n(\alpha) < n+1 .
\ee
In the absence of a defect, $\kp_n(0) = n+\frac{1}{2}$ is half-integer, thus reproducing the CdGM equidistant spectrum \eqref{CdGM-spectrum}. 

The main feature of Eqs.\ \eqref{eq-spectrum-E} and \eqref{eq-spectrum} is the opening of the gap in the excitation spectrum, which grows with the strength of the defect. The exact expression for the gap is given by
\be
\label{Eg-gen}
  E_g(\alpha,0) = \omega_0\sqrt{\kp_0^2(\alpha) + \alpha^2} .
\ee
According to Eq.\ \eqref{kn-window}, the lowest positive momentum $k_0(\alpha)$ cannot exceed 1. Therefore we obtain a linear scaling of the gap in the limit $\alpha\gg1$:
\be
  E_g(\alpha,0) = \omega_0\left[\alpha + O(\alpha^{-1})\right] .
\ee
This linear gap growth can be distinctly seen in Fig.\ \ref{F:alpha_levels}, where we plot the spectrum as a function of the defect strength $\alpha$.

\subsubsection{Coarse-grained density of states}

The opening of the gap is accompanied by redistribution of many levels. In the limit $\alpha\gg1$, approximately $\alpha$ states are strongly perturbed by the defect, and Fig.\ \ref{F:alpha_levels} demonstrates ``level crowding'' at $E$ above $E_g$. This effect can be described by the coarse-grained density of states $\rho(E)$, which is observed by replacing summation over the states by integration over $k$ [justified by the localization property \eqref{kn-window}]. As a result, the density of states takes a BCS-type form:
\be
\label{DOS}
  \rho(E) 
  = 
  \frac{1}{\omega_0}
  \re \frac{E}{\sqrt{E^2-E_g^2}} ,
\ee
(see inset to Fig.\ \ref{F:alpha_levels}).

We emphasize that Eq.\ \eqref{DOS} describes gap formation for the states of the chiral branch, and its magnitude $E_g\approx\omega_0\alpha$ is assumed to be much smaller than the bulk superconducting gap $\Delta$.

\subsubsection{Eigenfunctions}

The procedure described in Sec.\ \ref{shred_eq_text} allows us to immediately write an expression for the eigenfunctions. Working for simplicity with positive-energy states ($k_n>0$) and using Eq.\ \eqref{Psi(x)}, we obtain the wave function as a combination of two counterpropagating waves:
\be
\label{wf}
  \psi_n(x)
  =
  \psi_n(0) \times
  \begin{cases}
    C_1 e^{ik_nx}
  + C_2 e^{-ik_nx} ,
    & 0<x<\pi ,
    \\
    C_1^* e^{ik_nx}
  + C_2^* e^{-ik_nx} ,
    & -\pi<x<0 ,
  \end{cases}
\ee
with the coefficients
\be
\label{C1C2}
  C_1 = \frac{k_n+E_n/\omega_0-i\alpha}{2k_n} 
, \qquad
  C_2 = \frac{k_n-E_n/\omega_0+i\alpha}{2k_n} .
\ee
The value of the overall factor $\psi_n(0)$ should be determined from the normalization condition
$\int_{-\pi}^{\pi}|\psi_n(x)|^2 \, dx/2\pi = 1$. 
Using the spectral equation \eqref{eq-spectrum} and assuming $\psi_n(0)$ positive, we obtain
\be
\label{psi0}
  \psi_n(0)
  =
  \frac{k_n}{\sqrt{(E_n/\omega_0)^2+\alpha/\pi}} 
\ee
(the last term in the denominator can typically be safely neglected). One can show that the wave function acquires a phase shift $\pi(n+\frac{1}{2})$ when $x$ is increased by $\pi$:
\be
\label{psi:x+pi}
  \psi_n(x+\pi)=i(-1)^n\psi_n(x) ,
\ee
generalizing the same property in the clean case [with plane-wave functions $\psi_n(x)=e^{i(n+1/2)x}$] to arbitrary values of $\alpha$.

\begin{figure}
\includegraphics{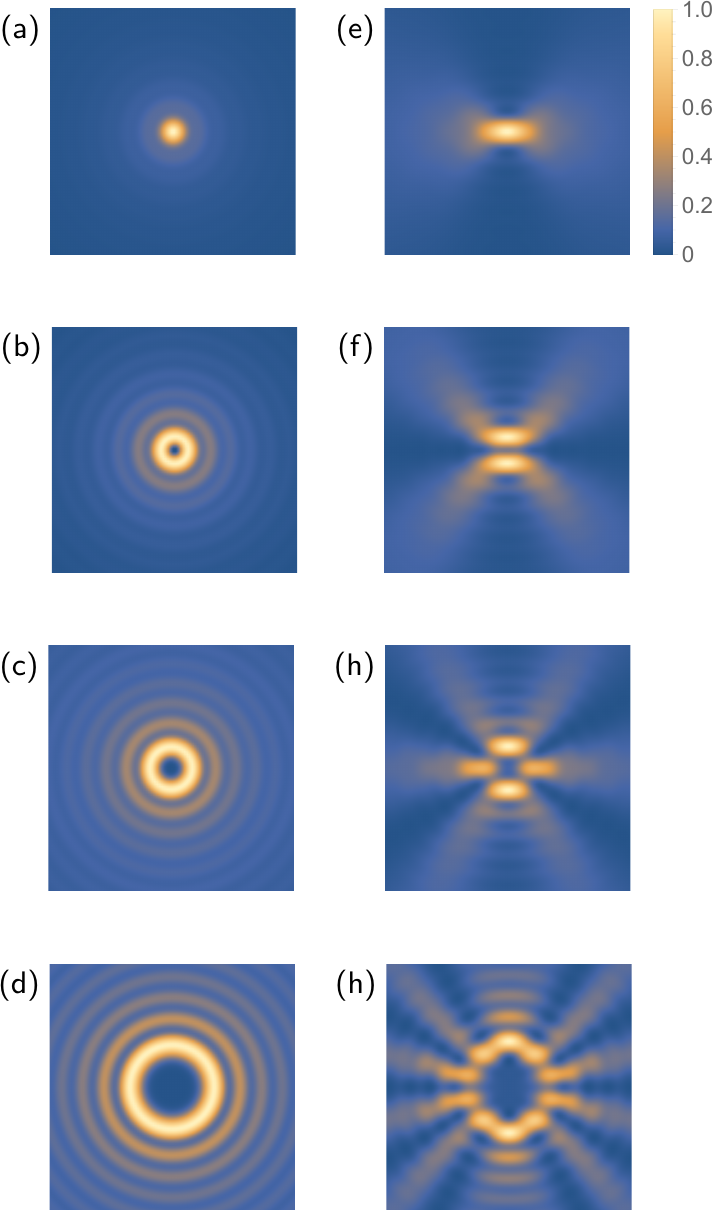}
\caption{Quasiparticle densities $P_n(r_1, r_2)$ for low-energy states $n = 0$, $1$, $2$, and $5$. Left: no defects, right: vertically-oriented linear defect with $\alpha=20$ passing through the vortex center. Shown is the region with $-20<(k_Fr_1,k_Fr_2)<20$.}
\label{F:wf}
\end{figure}

The two-wave structure of the eigenfunction \eqref{wf} has a simple physical interpretation. In a clean vortex, a low-energy quasiparticle trapped in its core exhibits Andreev precession \cite{Kopnin-book}. In the angular representation that corresponds to the plane wave $e^{ikx}$. A linear defect passing through the vortex center can normal-reflect a quasiparticle changing the sign of its momentum: $k\to-k$. Interference of such reflected waves is responsible for the formation of the states in the presence of the defect.

With the obtained structure of wave functions. one can easily compute the overlap of the $n$th eigenstate in the presence of the defect, $|n\rangle$, with the momentum eigenstate $|\mu\rangle$ for a clean vortex:
\be
\label{overlap}
  \corr{\mu|n}
  =
  \frac{[1+(-1)^{n+\mu-1/2}] \alpha  k_n}
  {\pi (k_n^2-\mu^2) \sqrt{(E_n/\omega_0)^2+\alpha/\pi}} .
\ee
The state $|n\rangle$ has the largest overlap with the clean state with the same ordinal number $n$ and momentum $\mu_n=n+\frac{1}{2}$. One can see that this overlap is always large, bounded by $\corr{\mu_n|n}>2/\pi$. For the lowest-energy state, $\lim_{\alpha\to\infty}\corr{\mu_0|0}=8/3\pi$. It means that for all $\alpha$ the exact state $|n\rangle$ is very close to the corresponding clean state $|\mu_n\rangle$, with a small admixture of satellites.

It is instructive to visualize reogranization of the quasiparticle states induced by the defect by looking at their wave functions $\Psi_n(\br)$ in real space. The wave function of the state $|n\rangle$ is obtained by expanding over the clean chiral basis:
\be
\label{psi-real-space}
  \Psi_n(\br) 
  = 
  \sum_\mu 
  \corr{\mu|n} \Psi_\mu(\br) ,
\ee
where the overlap is given by Eq.\ \eqref{overlap}.
Figure \ref{F:wf} shows the quasiparticle density profiles for low-energy states with $n = 0$, $1$, $2$, and $5$ without a defect (left column) and in the presence of a (vertical) defect with $\alpha=20$ (right column). The quasiparticle density $P_n(\br) = \Psi_n^\dagger(\br) \Psi_n(\br)$ shows the probability of finding an excitation (either its electron or hole component) at a given point. Note that $P_n(\br)$ is different from the charge density $\Psi_n^\dagger(\br) \tau_3 \Psi_n(\br)$, which weights electron and hole contributions with different signs ($\tau_3$ is the Pauli matrix in the Nambu space).

In Fig.\ \ref{F:wf} we see that a linear defect breaks the axial symmetry of $P_n(\br)$, which becomes corrugated in the angular direction. Note however that such a rough feature of the radial structure of $P_n(\br)$ as the peak at $k_Fr\sim n$, where $n$ is the state ordinal number, turns out to be stable.

\subsection{Vortex at a distance $b$ from the defect}
\label{SS:b>0}

In this case, the general approach described in Sec.\ \ref{shred_eq_text} is also formally applicable, but since the matrix $M_E$ in Eq.\ \eqref{M-def} contains now all three Pauli matrices, the transfer matrix $S_E$ given by the $T$\,exponent \eqref{S-T} cannot be evaluated in a closed form and the spectrum cannot be obtained analytically.

\begin{figure}
\centerline{\includegraphics[width=0.95\columnwidth]{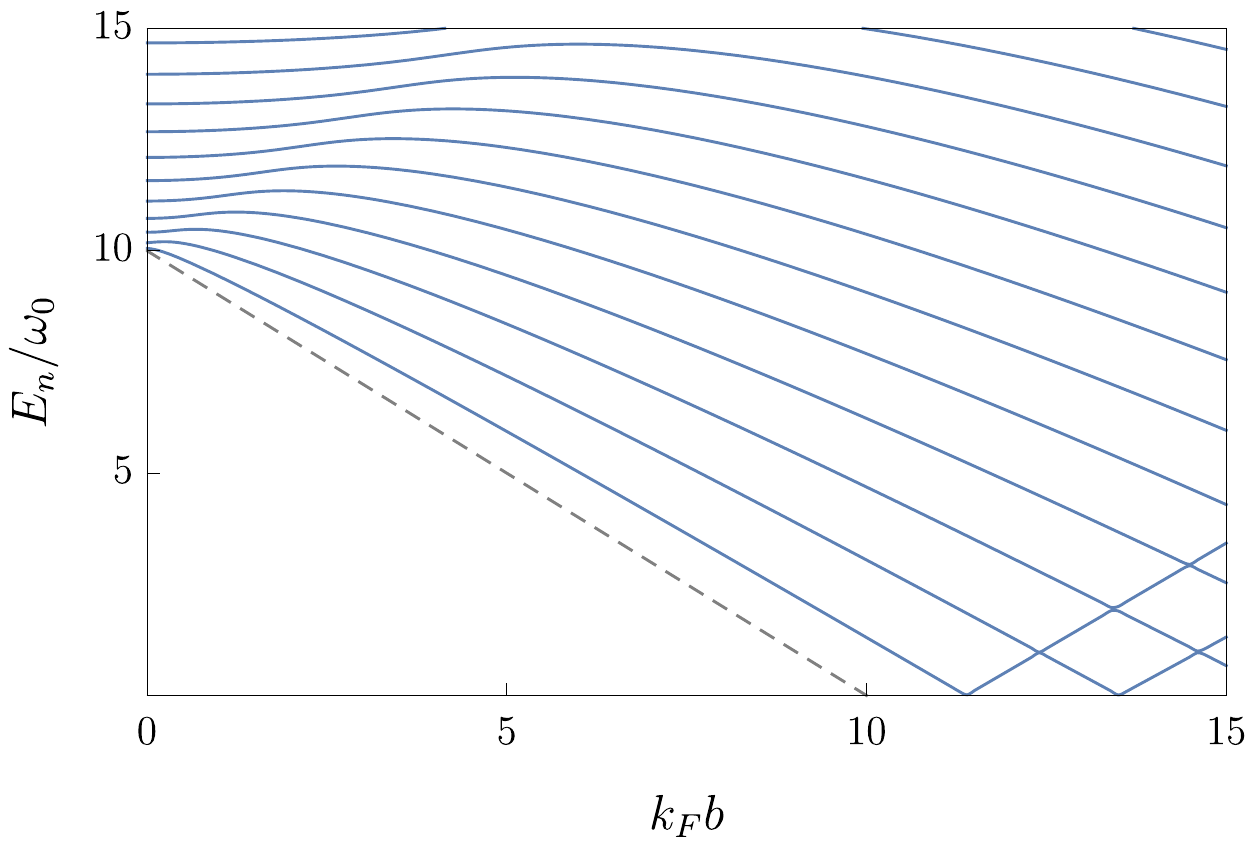}}
\caption{Positive energy levels vs.\ the distance $b$ between the vortex center and the linear defect of strength $\alpha = 10$.}
\label{F:levels_b}
\end{figure}

Numerical results for the spectrum evolution as a function of the parameter $k_Fb$ at fixed $\alpha=10$ are represented in Fig.\ \ref{F:levels_b}. The minigap is maximal, $E_g\approx\alpha$, when the vortex is located right at the defect, decreasing with the increase of $b$ in a nearly linear fashion until it closes approximately at $k_Fb=\alpha$.

The phenomenon of the minigap reduction illustrated in Fig.\ \ref{F:levels_b} can be explained analytically in the limiting case of $(k_F b,\,\alpha) \gg 1$. In this case it is more convenient not to use the spinor representation \eqref{shred_b_psi}, but to reduce the original quasi-local first order differential equation \eqref{L-eq} to the local second-order differential equation. Before doing that, we gauge out the phase factor by writing $\psi(x) = e^{ik_Fb\sin{x}}g(x)$ and then obtain the Schr\"odinger equation for the function $g(x)$ at the interval $0\leq x\leq\pi$:
\begin{equation}\label{g_x}
  - g''(x) + U(x) g(x) = 0
\end{equation}
with the potential
\begin{equation}
\label{U-def}
  U(x) 
  = 
  \alpha^2
  \left[1 - (t\cos{x} - \ve)^2\right] 
  +
  i\alpha t \sin{x}
\end{equation}
and the boundary conditions
\begin{subequations}
\label{g_bc}
\begin{gather}
  g'(0) = \alpha[1-i(t - \ve)] g(0) ,
\\
  g'(\pi) = - \alpha[1-i(t + \ve)] g(\pi) ,
\end{gather}
\end{subequations}
following from $2\pi$ antiperiodicity of $g(x)$.
Here we introduced the dimensionless parameters $t = k_F b/\alpha$ and $\ve = E/\alpha\omega_0$.
Equation \eqref{g_x} describes quantum-mechanical motion of a particle of mass $m = \frac{1}{2}$ in the potential $U(x)$ \emph{at zero energy}. The spectrum of the original problem ($\eps$) is determined by the requirement that such a zero-energy state exists.

In the limit of large $\alpha$, the last term in the potential \eqref{U-def} can be neglected:
\begin{equation}
  U(x) = \alpha^2 \left[1 - (t\cos{x} - \ve)^2\right] ,
\quad
  \alpha\to\infty ,
\end{equation}
and the boundary conditions \eqref{g_bc} dictate vanishing of $g$: $g(0)=g(\pi)=0$.
The necessary condition for the existence of a zero-energy state in the potential $U(x)$ is evidently $\min_x U(x)<0$. Right at the defect, it gives $\eps>1$, defining the position of the minigap $E_g\approx\alpha$. At finite $t$, the potential $U(x)$ becomes $x$-dependent, with the minimal value achieved at $x=\pi$, so we expand around it:
\be
\label{U(x)harm}
  U(x) 
  \approx 
  \alpha^2 \left[1 - (t+\ve)^2 + t(t+\ve)(\pi-x)^2 \right] .
\ee
Vanishing of $\min_x U(x)$ implies $\eps>1-t$, thus defining the leading $b$ dependence of the minigap: $\ve_g=1-t$. This quasi-classical estimate can be improved by taking into account quantum motion that requires $\min_xU(x)<0$ for the zero-energy ground state to exist. In the limit $\alpha\to\infty$, the potential $U(x)$ is sharp and the size of the ground state is small, justifying the expansion \eqref{U(x)harm}. Hence, we have a harmonic oscillator problem with the frequency $\omega=2\alpha\sqrt{t}$ (our mass is $\frac{1}{2}$).
However due to the rigid-wall boundary condition, $g(\pi)=0$, the ground state energy is not $\omega/2$ but rather $3\omega/2$. This gives $\ve_g=1-t+3\omega/4\alpha^2$, improving the above quasiclassical estimate estimate. In dimensional units,
\be
\label{Egb}
  \frac{E_g}{\omega_0} \approx \alpha - k_F b 
  + \frac{3}{2} \sqrt{\frac{k_Fb}{\alpha}} .
\ee
This formula perfectly explains why the minigap goes slightly above the dashed line $\alpha - k_F b$ in Fig.\ \ref{F:levels_b} and closes approximately at $k_Fb=\alpha+\frac{3}{2}$.

\begin{figure}
\centering
\includegraphics{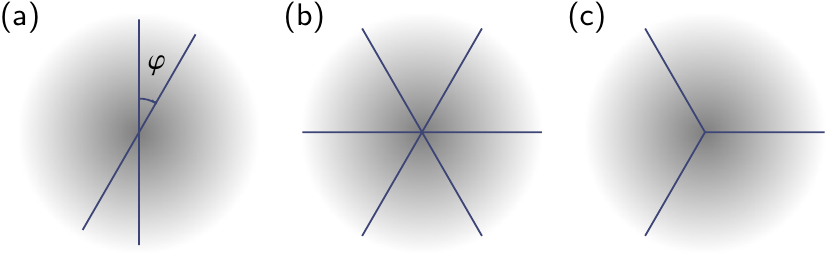}
\caption{Examples of linear defects intersecting at the vortex center: (a) two lines crossing at an angle $\vp$, (b) the most symmetric  arrangement with $n=3$ lines, (c) three-ray configuration.}
\label{F:few_lines}
\end{figure}

\section{Intersecting linear defects: commensurability effect}
\label{S:multiple}

In the previous section we have considered the case of a single linear defect crossing the vortex core. Now we are going to generalize the problem and consider a number of different configurations with several linear defects crossing at the vortex center (see Fig.\ \ref{F:few_lines}). This setup corresponds to a vortex sitting on the border of three or more grains in a granular medium.

\subsection{Two intersecting lines}
\label{two_lines}

We start with the simplest case of two linear defects intersecting at an angle of $\vp$ at the vortex center, 
as shown in Fig.\ \ref{F:few_lines}(a).
In order to construct the Hamiltonian we need to know matrix elements from both lines. The matrix element $V_{\mu\nu}^{(0)}$ for the vertical line is given by Eq.\ \eqref{V0-munu}.
One can easily verify that rotation of the line by an angle $\vp$ results in the appearance of a phase factor: \mbox{$V_{\mu\nu}^{(\vp)} = V_{\mu\nu}^{(0)} e^{i(\mu - \nu)\vp}$}. In the dual angular representation, this translates to the argument shift: $V^{(\vp)}(x,y) = V^{(0)}(x + \vp, y + \vp)$. Using Eq.\ \eqref{Vxy_b} with $b=0$, we obtain
\be
V^{(\vp)}(x,y) =i\alpha \omega_0 s(x+\vp) \times 2\pi\delta(x + y + 2\vp).
\ee
where here $s(x)=\sign(\sin x)$.

As a result, the two-line version of the Schr\"odinger equation \eqref{shred_b_psi} takes the form:
\begin{multline}
\label{Schr-eq-2}
-i\partial_x\psi(x) + i\alpha_1 s(x) \psi(-x)
\\{}
+i\alpha_2 s(x+\vp)\psi(-x-2\vp) 
= (E/\omega_0) \psi(x) ,
\end{multline}
where we assumed different defect strengths for generality.
The main difference from the original single-line problem \eqref{shred_b_psi}, which allowed for a local representation at the expense of introducing a two-component spinor \eqref{spinor} made of $\psi(x)$ and $\psi(-x)$, is that for the two-line problem such an approach typically fails. The reason is multiple reflections from the two lines that couple wave functions at the following arguments:
\be
\label{mirror}
  \pm x, \; \pm(x-2\vp), \; \pm(x-4\vp), \; \dots
\ee
Whether and where this sequence terminates (mod $2\pi$) depends on commensurability of $\vp$ and $\pi$:
\begin{itemize}
\item If $\vp=(m/n)\pi$ is a rational fraction of $\pi$ (coprime $m$ and $n$) then the set \eqref{mirror} contains $2n$ elements and one can introduce a $2n$-component vector $\Psi$ made of $\psi$ taken at the corresponding arguments. In terms of $\Psi$, Eq.\ \eqref{Schr-eq-2} becomes local and should be solved at the interval $x \in [0,\pi/n]$. That can be done as described in Sec.\ \ref{shred_eq_text}.
\item If $\vp$ is an irrational fraction of $\pi$ then Eq.\ \eqref{Schr-eq-2} cannot be brought to a local form.
\end{itemize}

The simplest is the case of two perpendicular lines ($\vp=\pi/2$), when the Schr\"odinger equation reduces to a local form in terms of a 4-component vector $\Psi$. Its analysis performed in Appendix~\ref{cross} allows us to determine the spectrum at arbitrary $\alpha_1$ and $\alpha_2$ by solving the transcendental equation \eqref{lambda-eq} similar to Eq.\ \eqref{eq-spectrum}.
In the limit of strong defects, the minigap is given by $E_g=\omega_0\sqrt{\alpha_1^2+\alpha_2^2}$.

Although the same analysis can be formally done for any rational $\vp/\pi=m/n$, the transcendental spectral equation becomes more and more complicated. However, the asymptotic behavior of the minigap at large $\alpha_i$ can be obtained in a closed form, as it does not rely on the knowledge of momenta. We calculate it in Appendix \ref{AA:minigap-asympt}, arriving at the following asymptotic expression:
\be
\label{Eg-asympt}
  E_g = 
  \omega_0\sqrt{\alpha_1^2+\alpha_2^2-2\alpha_1\alpha_2\cos(\pi/n)} ,
\ee
which depends only on the denominator $n$ of $\vp/\pi$.

\begin{figure}
\includegraphics[width=\columnwidth]{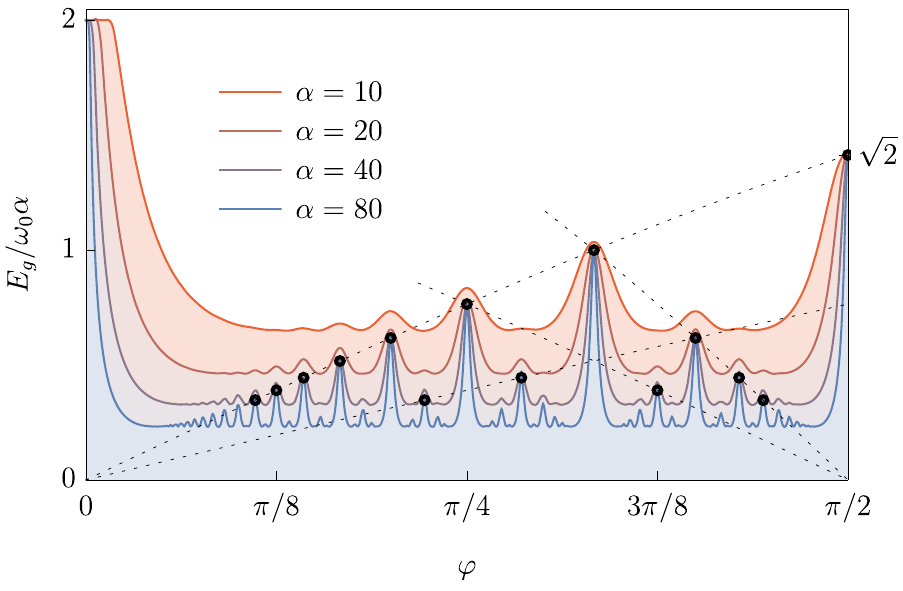}
\caption{Normalized minigap energy $E_g/\alpha\omega_0$ for two identical intersecting lines as a function of the crossing angle $\vp$ [see Fig.\ \ref{F:few_lines}(a)] for different values of the defect strength $\alpha$.}
\label{F:popcorn}
\end{figure}
 
For an arbitrary crossing angle $\vp$, Eq.\ \eqref{Schr-eq-2} should be solved numerically. In the symmetric case $\alpha_1=\alpha_2=\alpha$, the results are presented in Fig.\ \eqref{F:popcorn}, which shows the ratio $E_g/\alpha\omega_0$ as a function of $\vp$ for various values of the defect strength $\alpha$.
Figure \eqref{F:popcorn} has a number of remarkable features:
\begin{enumerate}
\item[(i)]
The appearance of peaks at commensurate angles $\vp=(m/n)\pi$, which become more pronounced and sharper with increasing $\alpha$.
\item[(ii)]
The presence of a nearly constant nonzero background 
for ``not very rational'' angles well described by the empirical formula $E_g^\text{bg}/\alpha\omega_0\approx 2/\sqrt{\alpha}$.
\item[(iii)]
Independence of the peak height (once resolved) on the numerator $m$.
\end{enumerate}

This picture is consistent with the minigap asymptotics \eqref{Eg-asympt} with the the following limiting dependence:
\be
\label{eq-popcorn}
  \lim_{\alpha\to\infty} \frac{E_g(\vp)}{\alpha\omega_0}
  =
  2 \sin\left( \frac\pi2 \popcorn\frac\vp\pi \right) .
\ee
Here $\popcorn(x)$ is the Thomae's function (also referred to as the popcorn function), which takes zero value for irrational $x$ and $1/n$ for rational $x=m/n$ (with $m$ and $n$ coprime). Dashed lines in Fig.\ \ref{F:popcorn} are drawn with the help of Eq.\ \eqref{eq-popcorn} along a number of the principal peaks  of the Thomae's function.

Equation \eqref{eq-popcorn} predicts that peak heights at commensurate angles, $E_g[(m/n)\pi]\sim\alpha\omega_0/n$, grow linearly with the defect strength $\alpha$. However due to the presence of a finite background $E_g^\text{bg}\sim\sqrt{\alpha}\omega_0$, only peaks with $n\lesssim\sqrt\alpha$ can be actually resolved. We emphasize that the energy scale $E_g^\text{bg}\approx2\sqrt{\alpha}\omega_0$ also grows with $\alpha$, but as a square root (this growth can be accidentally overlooked in Fig.\ \ref{F:popcorn}, where $E_g$ is normalized by $\alpha$). Hence the value of $2\sqrt{\alpha}\omega_0$ provides a lower bound for the minigap at arbitrary angles.

In order to qualitatively understand the origin of the background minigap energy $E_g^\text{bg}$ and its $\alpha$ dependence, we recall that the minigap at a commensurate angle $\vp=(m/n)\pi$ can be written as $E_g=\sqrt{\lambda_\text{min}^2+k_0^2} \, \omega_0$. In physical terms, $\lambda_\text{min}$ and $k_0$ (which both are functions of $\alpha$ and $n$) provide the contributions of the potential and kinetic energy to the minigap, respectively. Here $\lambda_\text{min}=\lambda(q_1)$ is the minimal positive eigenvalue of the matrix $iR$ in Appendix \ref{AA:minigap-asympt}, which determines the gap asymptotics via $E_g=\lambda_\text{min}\omega_0$ [cf.\ Eq.\ \eqref{Eg-asympt}]. The parameter $k_0$ is the first positive solution of the transcendental spectral equation, in an explicit form given by Eq.\ \eqref{eq-spectrum} for one line and \eqref{lambda-eq} for two perpendicular lines. One can easily show that $\lim_{\alpha\to\infty}k_0(\alpha)=1$ for one line and $\lim_{\alpha\to\infty}k_0(\alpha)=2$ for two perpendicular lines. Thus we see that $k_0$ grows with the denominator of $\vp/\pi$ and it is natural to assume that $\lim_{\alpha\to\infty}k_0(\alpha)\sim n$. Now comparing the decreasing potential-energy contribution $\lambda_\text{min}=\lambda(q_1)\sim\alpha/n$ with the increasing kinetic-energy contribution $k_0\sim n$ we obtain that they become comparable at $n\sim\sqrt\alpha$, when $E_g$ just coincides with obtained background minigap level $E_g^\text{bg}$. We believe the above arguments qualitatively explain the relevance of the kinetic energy in the background minigap formation and provide an estimate for its magnitude.

\subsection{Several intersecting lines}

For completeness, we also discuss the case of the most symmetric configuration of $n$ identical lines of strength $\alpha$ intersecting at the vortex center at the angle of $\vp=\pi/n$ [see Fig.\ \ref{F:few_lines}(b)]. The Schr\"odinger equation, which now takes into account scattering from $n$ lines, may be brought to a local form by introducing a $2n$-component vector $\Psi$ in the same manner as described above. The asymptotic minigap behavior at $\alpha\gg1$ can be obtained by the method developed in Appendix \ref{AA:minigap-asympt}. After some algebra, we obtain
\be
  E_g
  =
  \alpha\omega_0
  \times
  \begin{cases}
    1 , & \text{$n$ odd} , \\
    1/\cos(\pi/2n) , & \text{$n$ even} . \\
  \end{cases}
\ee

Surprisingly, the minigap remains of the order of $\alpha\omega_0$ regardless of the number of intersecting lines. The fact that it does not scale with a naive estimate of ``the overall defect strength'' $n\alpha$ is a consequence of destructive interference of waves multiply scattered from different defects. At the same time, periodicity of the structure ensures that the minigap is not destroyed completely but remains finite with $E_g\approx\alpha\omega_0$.
Addition of any imperfections would spoil this picture and suppress $E_g$, presumably not completely but at least to the level of $\sqrt\alpha\omega_0$.

\subsection{Three rays}

\begin{figure}
\centerline{\includegraphics[width=0.95\columnwidth]{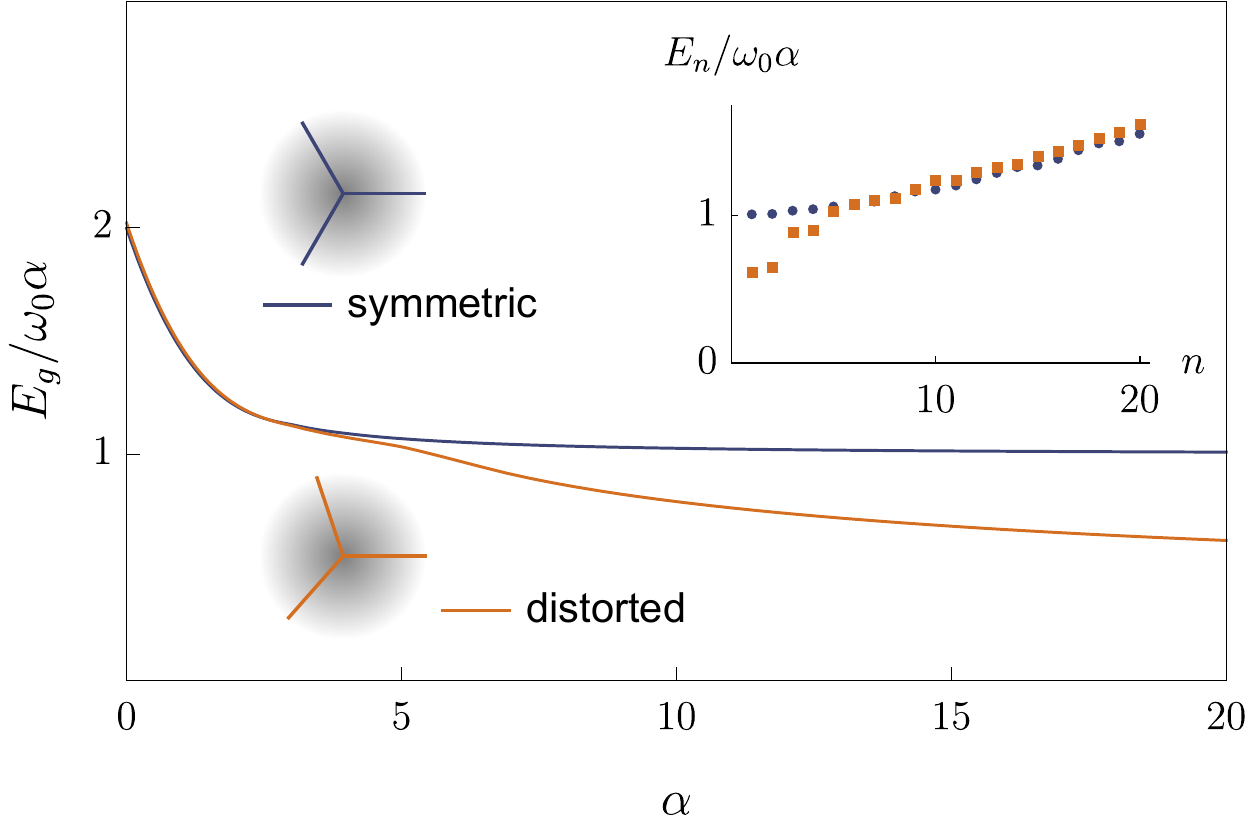}}
\caption{Normalized minigap $E_g/\alpha\omega_0$ in the three-ray configuration [see Fig.\ \ref{F:few_lines}(c)] as a function of $\alpha$. Upper line: symmetric star with all angles $2\pi/3$; lower line: distorted star with one ray rotated by 0.2 rad.
Inset: energy levels $E_n$ vs.\ $n$ in the symmetric (blue points) and distorted (orange 
boxes) configurations at $\alpha = 20$.}
\label{star}
\end{figure}

Finally we address the three-ray configuration depicted in Fig.\ \ref{F:few_lines}(c), which mimics a contact of three grains. All half-line defects are assumed to have the same strength per unit length, $\alpha$.
The vortex center is located at the rays' intersection point. The matrix elements of the ray defect $V(\br)\propto\delta(r_1)\theta(r_2)$ are different from those of the linear defect $V(\br)\propto\delta(r_1)$ and are calculated in Appendix \ref{AA:half-line}. In the angular representation they become essentially non-local [Eq.\ \eqref{Vxy-ray}]. This makes it impossible to obtain an analytic solution, and we perform numeric analysis in the $\mu$ representation [Eq.\ \eqref{Vmunu-ray}]. 

We consider two configurations, symmetric with the angles between the rays equal $2\pi/3$ and distorted with the angles $2\pi/3$ and $2\pi/3\pm0.2$, and study the spectrum as a function of the defect strength $\alpha$. The results for the corresponding minigaps are shown in Fig.\ \ref{star}. In the symmetric case we obtain a linear scaling $E_g \approx \alpha\omega_0$, whereas the minigap in the distorted geometry is suppressed, growing approximately as $\sqrt\alpha$
for large $\alpha$.

The fact that already a small rotation of one ray by an angle 0.2 leads to a significant minigap suppression, which becomes more pronounced in the limit $\alpha\to\infty$, is fully consistent with the commensurability effect for two intersecting lines discussed in Sec.\ \eqref{two_lines}: For rational angles with significantly small denominators, the minigap $E_g\sim\alpha\omega_0$. Otherwise, destructive interference from different lines suppresses it to a background level of $E_g^\text{bg}\sim\sqrt\alpha\omega_0$.

Inset to Fig.\ \ref{star} shows the spectrum $E_n$ (its positive part) vs.\ ordinal number $n$ for the symmetric (blue circles) and distorted (orange boxes) configurations at $\alpha=20$.
Although the energy $E_0$ of the lowest level (and thus the minigap) in the distorted case is already significantly reduced compared to $\alpha\omega_0$, only few levels visibly change their position compared to the symmetric case. Therefore, the coarse-grained density of states will still have a BCS singularity \eqref{DOS} at $E=\alpha\omega_0$, with a small fraction of ``subgap states'' with $E<\alpha\omega_0$.

\section{Periodic structures of defects}
\label{S:Periodic}

\subsection{Square arrays of linear defects}

\begin{figure}
\centering
\includegraphics[width=0.95\columnwidth]{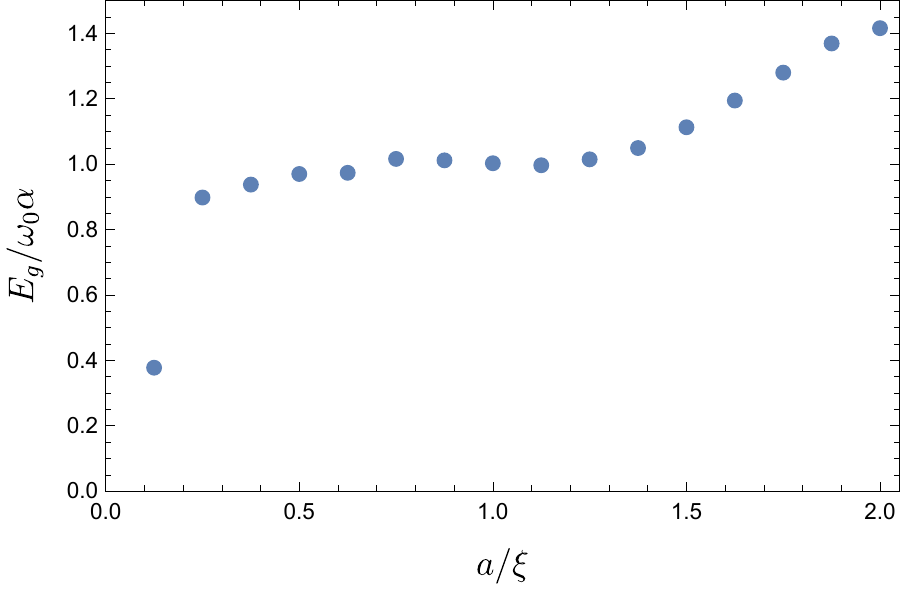}
\caption{Minigap in the case of the square lattice of linear defects as a function of the ratio of the lattice period $a$ to the superconducting coherence length $\xi$. Defect strength $\alpha = 20$, $k_F\xi = 200$.}
\label{lattice}
\end{figure}

Keeping in mind connection to granular systems, we consider here a square grid of potential lines with the period $a$ and the vortex center at one of the grid nodes. One can ask whether destructive interference from different lines can suppress and/or totally destroy the minigap.

The matrix elements of the grid $V_{\mu\nu}$ can be calculated following the procedure described in Secs.\ \ref{SS:matelems} and \ref{two_lines}. 
The resulting minigap obtained numerically is shown in Fig.\ \ref{lattice} (for $\alpha=20$ and $k_F\xi=200$). It reaches its asymptotic value $E_g=\sqrt{2}\alpha\omega_0$ [see Eq.\ \eqref{Eg-asympt}] at $a\gg\xi$ and gradually decreases with the decrease of the lattice period. Nevertheless, $E_g$ remains of the order of $\alpha\omega_0$ in a broad range of $a/\xi$, with a visible suppression at $a<\xi/4$.
Thus, we conclude that the phenomenon of the minigap opening is observed for periodic structures as well, provided that the lattice period much exceeds the Fermi wavelength.

\subsection{Approximating defect line by point defects}

It was mentioned in the Introduction that the minigap does not appear in the presence of point-like impurities. On the other hand, a linear defect can be formally considered as a dense pack of weak point-like impurities. To study a crossover from the linear defect to such an array of point-like defects, we consider the following model potential:
\be
\label{V-lin-point}
  V(\br) 
  = 
  \frac{\hbar^2\varkappa a}{m}
  \sum_n \delta(r_1) \delta(r_2-na) .
\ee
When its period $a$ vanishes, it reproduces the linear defect potential \eqref{V-ini}.

\begin{figure}
\centerline{\includegraphics[width=0.95\columnwidth]{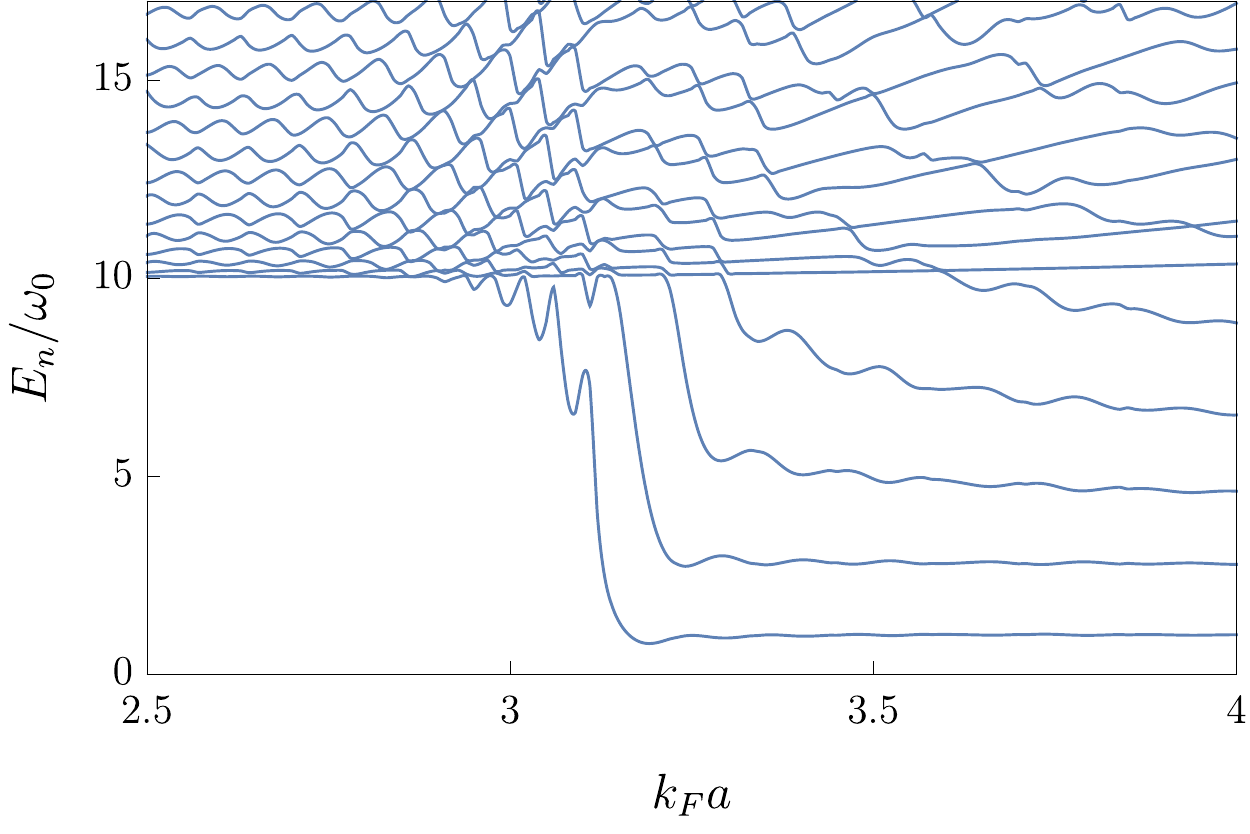}}
\caption{Positive energy levels for a 1D lattice of point-like impurities \eqref{V-lin-point} as a function of the lattice period $a$. Defect strength $\alpha = 10$.}
\label{fence}
\end{figure}

The energy spectrum obtained numerically is shown in Fig.\ \ref{fence}. It demonstrates a distinct transition between the gapped and gapless phases taking place at $k_F a = \pi$. 
One can provide the following qualitative explanation of this phenomenon.
For a point-like defect, an incident wave is scattered in all directions, whereas the linear defect acts like a mirror. If $k_F a < \pi$, the waves reflected from adjacent point-like defects are coherent and interfere with each other: the lattice acts like a diffraction grating. Otherwise, the waves reflected from adjacent point defects are incoherent, and no minigap opens.

\section{Region $\alpha\gg\sqrt{k_F\xi}$: Subgap states and ``soft gap'' of Ref.\ \cite{Melnikov2020}}
\label{S:NN}

Having analyzed various defect configurations, now we come back and revisit the simplest case of a single defect line passing through the vortex center. On one hand, there is a sufficiently transparent derivation of the matrix element $V(x,y)$ leading to the delta-function expression \eqref{Vxy_b}. The resulting quantum mechanics studied in Sec.\ \ref{zero_distance} is rather simple and does not contain any knowledge of the parameter $k_F\xi$. On the other hand, Fig.\ \ref{F:levels-M} obtained using the exact matrix elements $V_{\mu\nu}$, in accordance with Ref.\ \cite{Melnikov2020}, clearly demonstrates the existence of a different regime at $\alpha\gtrsim\sqrt{k_F\xi}$, with a number of states sequentially splitting off the majority of gapped states. In the language of quasiclassical trajectory analysis of Ref.\ \cite{Melnikov2020}, appearance of those states is associated with a topological transition in the phase space. They correspond to special trajectories, which do not precess but are aligned along the defect. Below we discuss how this effect can be understood in terms of the quantum mechanics developed in Sec.\ \ref{zero_distance}.

A mechanism responsible for the breakdown of a simple picture discussed in Sec.\ \ref{S:one-line} is smearing of the delta function $\delta(x+y)$ in Eq.\ \eqref{Vxy_b} for the matrix element $V(x,y)$.
This formula was obtained from the exact expression \eqref{V-xy-line-0} by neglecting the envelope factor $e^{-2K}$. If it is not neglected, the delta function in Eq.\ \eqref{delta-A9} will acquire a finite width of the order of $1/\xi$, and the delta function $\delta(x+y)$ in Eq.\ \eqref{Vxy_b} will be smeared by an $x$-dependent amount of $1/k_F\xi|\sin x|$. Since $k_F\xi\gg1$, this smearing is typically small except for very small angles $|x|\lesssim x_*$, where $x_*=1/\sqrt{k_F\xi}$. Hence, at $x,y\sim x_*$ the matrix element $V(x,y)$ should be considered as an integral kernel, while outside of this interval it can be approximated by the delta-function form \eqref{Vxy_b}.

\begin{figure}
\centering
\includegraphics{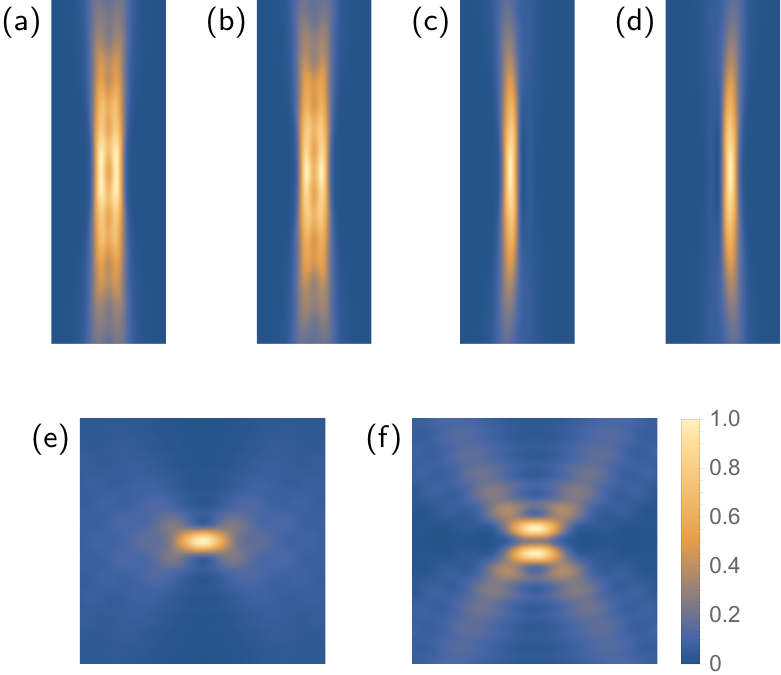}
\caption{Quasiparticle densities $P_n(r_1, r_2)$ for the states shown in Fig.\ \ref{F:levels-M} at $\alpha=30$: 
(a), (b) lowest pair of subgap states [which can be considered as symmetric and antisymmetric combinations of the states localized to the left (c) and to the right (d) of the linear defect], (e), (f) next two states. Shown is the region with $-100<k_Fr_1<100$ and $-300<k_Fr_2<300$ for (a)--(d), and $-20<(k_Fr_1,k_Fr_2)<20$ for (e) and (f). Vortex size is specified by $k_F\xi=200$.}
\label{F:wf2}
\end{figure}

Sequential splitting off the subgap state taking place at $\alpha\gtrsim\sqrt{k_F\xi}$ corresponds to the appearance of new types of eigenstates of the Hamiltonian 
$-i\partial_x+V$,
which are localized either at $x\sim x_*$ or at $|x-\pi|\sim x_*$. The fact that their energy is smaller than $\alpha\omega_0$ indicates that the wave functions of these states have an imaginary momentum and decay exponentially away from the mentioned vicinities of 0 and $\pi$. The exponentially small energy difference between these states seen in the inset to Fig.\ \ref{F:levels-M} is a consequence of exponentially weak hybridization of the states at 0 and at $\pi$. 

Figures\ \ref{F:wf2}(a) and\ \ref{F:wf2}(b) provide the snapshots of the quasiparticle density $P_n(\br)$ for the pair of lowest-energy subgap states at $\alpha=30$. These states are clearly aligned along the linear defect and can be considered as symmetric and antisymmetric combinations of the states localized to the left and to the right from the defect, as shown Figs.\ \ref{F:wf2}(c) and\ \ref{F:wf2}(d). At the same time, Figs.\ \ref{F:wf2}(e) and\ \ref{F:wf2}(f) show the quasiparticle density for the third and fourth positive-energy states at $\alpha=30$. Note that they look pretty similar to the pair of lowest-energy states in Figs.\ \ref{F:wf}(e) and\ \ref{F:wf}(f). It means that the behavior of the majority of the states is almost insensitive to the existence of the subgap states and still can be described by the simple theory developed in Sec.\ \ref{S:one-line}. Such a situation is consistent with separation of the phase space into two regions reported in Ref.\ \cite{Melnikov2020}.

To conclude this section, we emphasize that though our fully microscopic quantum-mechanical analysis confirmed the existence of subgap states propagating along the defect at $\alpha\gtrsim\sqrt{k_F\xi}$, their description in terms of the developed formalism is rather complicated. We believe the trajectory-based approach \cite{Melnikov2020} is more suitable for this purpose.

\section{Effect of $\Delta(\br)$ distortion}\label{S:self_consistency}

In our consideration above, the profile of the order parameter, $\Delta(\br)$, was assumed to be unchanged by the linear defect. The same approximation was used in Ref.~\cite{Melnikov2020}.
However, since the order parameter should be determined self-consistently, modification of the quasiparticle spectrum does have an impact on $\Delta(\br)$ (in particular, it loses its axial symmetry). Deformation of the order parameter, in turn, affects the quasiparticle states, which therefore should be determined self-consistently. However, such a procedure is rather complicated and in a non-uniform vortex geometry is nearly intractable. Nevertheless, it often happens that a direct effect of a perturbation is more important that the accompanying effect of $\Delta(\br)$ modification. 

Assuming this is also the case for our problem with a linear defect, we can treat the effect of $\Delta(\br)$ distortion perturbatively. We take the quasiparticle states obtained above for the clean order parameter $\Delta_0(\br)$ and substitute them to the self-consistency equation to obtain the next iteration for the order parameter, $\Delta_1(\br)$. Then the difference $\delta\Delta(\br)=\Delta_1(\br)-\Delta_0(\br)$ is considered as a perturbation in the BdG equation \eqref{BdG} and the correction to the minigap $\Delta E_g$ is obtained.

This procedure is performed in Appendix \ref{A:self_consistency}. We obtain that for $\alpha>10$ the minigap shift nearly saturates at $\Delta E_g \approx 0.1 \times 4\pi g A^4/k_F^2$, where $g$ is the BCS coupling constant. Using a model dependence of the vortex order parameter, $\Delta(r) = \Delta_0 r/\sqrt{r^2 + \xi^2}$, we obtain for $\alpha>10$
\be
  \frac{\Delta E_g}{\omega_0} 
  \approx
  0.5 \, \nu_0 g ,
\ee
where $\nu_0$ is the normal-state DOS at the Fermi level. In the BCS theory, the dimensionless coupling constant $\nu_0 g<1$. Since the minigap grows with $\alpha$ as $E_g=\alpha\omega_0$, we conclude that the distortion of the order parameter does not have a significant effect on the quasiparticle excitation spectrum.

\section{Discussion and conclusion}
\label{S:Conclusion}

In this publication, we report on extensive studies of the quasiparticle states localized in the pancake vortex in a clean 2D superconductor in the presence of one or several linear defects (that corresponds to a vortex in a layered 3D superconductor with planar defects).

In the configuration with one linear defect passing through the vortex center, we identify two different regimes separated by $\alpha\sim\sqrt{k_F\xi}$. For sufficiently weak defects with $\alpha\lesssim\sqrt{k_F\xi}
$, the spectral problem can be solved exactly. The spectrum is characterized by a minigap growing as $E_g=\alpha\omega_0$ for $\alpha\gg1$ and the BCS-type DOS \eqref{DOS} above the minigap.
The wave functions in this regime are localized in the angular momentum ($\mu$) representation and delocalized in the dual angular representation. In the quasiclassical description it means that a trajectory exhibits the standard Andreev precession weakly modulated by scattering off the linear defect.
For stronger defects with $\alpha\gtrsim\sqrt{k_F\xi}
$, the majority of states are still gapped with $E_g=\alpha\omega_0$; however, a number of subgap states emerges with $E<E_g$. These states are localized either to the left or to the right from the defect and hence are delocalized in the $\mu$ representation. The pairs of subgap states sequentially split off the bulk states with increasing $\alpha$, as shown Fig.\ \ref{F:levels-M}. Hence, for a single line we reproduce and confirm the prediction of Ref.\ \cite{Melnikov2020}.

We also analyzed reorganization of the chiral states in a number of more complicated geometries with linear defects. Here we assumed $\alpha\lesssim\sqrt{k_F\xi}$, such that complications due to formation of subgap states do not appear and the problem can be mapped onto a sufficiently simple quantum mechanics.

We considered a configuration with the vortex center located at a finite distance $b$ from the linear defect. Such a configuration cannot be realized at equilibrium since the vortex prefers to minimize its potential energy and chooses to sit right at the defect. However, such a situation can take place in the presence of a depinning force or as a dynamic state under microwave absorption. We find that the minigap decreases with the growth of $b$ and closes at $b\approx\alpha/k_F$.

A vortex pinned at the intersection of two linear defects demonstrates a peculiar commensurability effect, when the minigap essentially depends on how close is the angle $\vp$ between the defects to a rational of $\pi$. For perfect matching, the minigap $E_g\approx\alpha\omega_0$, while in the most frustrated case the minigap still exists but at a smaller background level of $E_g\approx\sqrt\alpha\omega_0$ (see Fig.\ \ref{F:popcorn}).

The phenomenon of the minigap opening survives in the presence of a periodic structure of linear defects, even if the period $a$ is smaller than the coherence length. For a square lattice of equal defects, the minigap $E_g\sim\alpha\omega_0$, unless the $a$ becomes comparable to the Fermi wavelength.

Having considered various types of linear defects, we conclude that the effect of gap opening is quite robust. This observation has obvious consequences for vortex behavior in a granular media. Vortices pinned at grain boundaries are expected to have a minigap typically scaling with the boundary strength $\alpha$ as $E_g\approx\alpha\omega_0$. This effect will manifest itself at low temperatures $T<E_g$ in the exponential suppression of heat capacity and flux-flow conductivity, as well as in a threshold behavior of optical conductivity.
Low-temperature anomalies in flux-flow conductivity have been recently reported in granular aluminum \cite{Plourde}. Although this material belongs to the dirty case, we believe that the qualitative conclusion on the gap opening in granular systems remains valid in the presence of disorder and therefore can explain the experimental finding of Ref.\ \cite{Plourde}.

Our analysis of the electronic states in the vortex core is applicable for highly transparent defects with the reflection coefficient $R \ll 1$, corresponding to the inequalities $\alpha \ll k_F\xi$ and $E_g \ll \Delta$. In this limit the defect perturbation effectively redistributes only $\alpha$ lowest states, without admixing the states of continuous spectrum. Therefore, we can access neither the Josephson vortex regime (realized in the tunneling limit with $T\ll1$) nor the crossover from the Abrikosov to the Josephson vortex. Nevertheless, one can argue that the subgap states localized along the defect are presumably important for the transition from the Abrikosov to the Josephson vortex \cite{JV} with increasing the defect strength $\alpha$. In particular, our conclusion that modification of the order parameter is not important for obtaining the minigap (see Sec.\ \ref{S:self_consistency}) might be modified if those gliding states are taken into account.

Finally, we mention that the developed theory can be easily generalized to the case of linear defects in clean $p$-wave superconductors. We expect the zero-energy Majorana bound state \cite{Dima-pwave} will survive gap opening and will facilitate transport across the gap.

\acknowledgments

We are grateful to Ya.\ V. Fominov, V. B. Geshkenbein, D. A. Ivanov, A. S. Melnikov and G. E. Volovik for stimulating discussions. This work was partially supported by the Russian Science Foundation under Grant No.\ 20-12-00361.

\appendix

\section{KERNEL $V(x,y)$ IN THE ANGULAR REPRESENTATION}
\label{V_x_y}

\subsection{General expression}
\label{AA:V-gen}

In the chiral basis, the matrix elements of a generic potential perturbation are given by Eq.\ \eqref{V_def}. Taking the wave functions from Eq.\ \eqref{wf} and tracing over the Nambu space, we obtain
\be
\label{V-munu}
  V_{\mu\nu} 
  = 
  A^2 \int d^2r \,
  e^{-2K(r)}
  e^{i(\nu-\mu)\vp}
  w_{\mu\nu}(k_Fr)
  V(\br) 
  ,
\ee
where
\be
  w_{\mu\nu}(z)
  =
  J_{\mu-1/2}(z)J_{\nu-1/2}(z) 
- J_{\mu+1/2}(z)J_{\nu+1/2}(z) .
\ee

Now we transform the matrix $V_{\mu\nu}$ to the angular representation according to Eq.\ \eqref{fourier}. Summation over momenta is done with the help of the Jakobi-Anger identity
\be
\label{Anger}
  e^{iz\sin\theta}
  =
  \sum_{n} J_n(z) e^{in\theta} ,
\ee
leading to
\be
   \sum_{\mu\nu}e^{i(\nu-\mu)\vp}
   w_{\mu\nu}(k_Fr)e^{ix\mu-iy\nu}  
  =
  2i\sin\frac{x-y}{2}
  e^{ik_FR_{xy}(\br)} ,
\ee
where
\be
\label{Rxy-def}
  R_{xy}(\br)
  =
  r[\sin(x-\vp) - \sin(y-\vp)] .
\ee
In terms of the Descartes coordinates, $r_1=r\cos\vp$ and $r_2=r\sin\vp$, $R(\br)$ is given by
\be
  R_{xy}(\br)
  =
  r_1(\sin x - \sin y)
+ r_2(\cos x - \cos y) .
\ee
Hence, the general expression for the kernel $V(x,y)$ valid for any potential $V(\br)$ takes the form
\be
\label{V-xy-gen}
  V(x,y)
  = 
  2i
  A^2 
  \sin\frac{x-y}{2}
  \int d^2r \,
  e^{-2K(r)}
  e^{ik_FR_{xy}(\br)} 
  V(\br) 
  .
\ee
The factor $\sin[(x-y)/2]$ reflects $2\pi$ antiperiodicity of wave functions in the angular representation [see Eq.\ \eqref{psi-2pi}].

\subsection{Kernel for the linear defect}
\label{AA:V-linear}

For the linear defect with $V(\br)$ given by Eq.\ \eqref{V-ini}, the coordinate $r_1$ coincides with $b$.
It then remains to integrate over the coordinate $r_2$ along the defect:
\begin{multline}
\label{V-xy-line-0}
  V(x,y)
  = 
  i \alpha \omega_0 k_F
  e^{ik_F b(\sin x - \sin y)}
  \sin\frac{x-y}{2}
\\{}
  \times
  \int_{-\infty}^\infty dr_2 \,
  e^{-2K\bigl(\sqrt{r_2^2+b^2}\bigr)}
  e^{ik_F r_2(\cos x - \cos y)} ,
\end{multline}
where we expressed the prefactor in terms of the dimensionless defect strength $\alpha$ introduced in Eq.\ \eqref{alpha_kappa}.

Expression \eqref{V-xy-line-0} is still an exact matrix element $V(x,y)$ for the linear defect, without any approximations. As far as we are interested in rearrangement of low-energy states by a not very strong defect ($\alpha\omega_0\ll\Delta$), one can further simplify $V(x,y)$. In this case, only clean states with momenta $\mu\sim\alpha$ are involved.
So we may replace the exponent $e^{-2K(r)}$, which decays exponentially at $r\sim\xi$, by 1. Then, the integral in the second line of Eq.\ \eqref{V-xy-line-0} produces the following delta function:
\be
\label{delta-A9}
  2\pi\delta[k_F(\cos{y} - \cos{x})])
  =
  \frac{2\pi[\delta(x - y) + \delta(x + y)]}{k_F|\sin x|} .
\ee
The first term in the right-hand side does not contribute due to vanishing of the factor $\sin[(x-y)/2]$ in Eq.\ \eqref{V-xy-line-0}, while the second term yields the matrix element \eqref{Vxy_b}.

\subsection{Kernel for the half-line defect}
\label{AA:half-line}

In this Appendix we calculate the matrix element in the angular representation, $V(x,y)$, for a half-line terminating at the vortex center and specified by the potential $V(\br)=({\hbar^2\varkappa}/{m})\delta(r_1)\theta(r_2)$, with $\theta(r_2)$ being the step function. The matrix element $V(x,y)$ can be written as $V^\text{line}(x,y)/2+\delta V(x,y)$, where $V^\text{line}(x,y)$ is the matrix element of the linear defect given by Eq.\ \eqref{Vxy_b} with $s(x)=\sign(\sin x)$, and the difference is defined as [cf.\ Eq.\ \eqref{V-xy-line-0}]
\begin{multline}
\label{half_line_xy}
  \delta V(x,y)
  = 
  \frac{i}{2} \alpha \omega_0 k_F
  \sin\frac{x-y}{2}
\\{}
  \times
  \int_{\infty}^\infty dr_2 \, \sign r_2 \,
  e^{-2K(r_2)}
  e^{ik_F r_2(\cos x - \cos y)} .
\end{multline}
Neglecting the factor $e^{-2K(r_2)}$ 
as it was done for the linear defect in Appendix \ref{AA:V-linear},
one gets 
$1/[k_F(\cos x - \cos y)]$
for the integral in the second line of Eq.\ \eqref{half_line_xy}. Hence we obtain the the matrix element of the half-line:
\be
\label{Vxy-ray}
  V(x,y)
  = 
  \frac{i
  \alpha
  \omega_0
  }{2}
  \left[
  s(x) 
  \times 2\pi \delta(x+y) 
  -
  \frac{1}{2\sin[(x+y)/2]} 
  \right] .
\ee
Making Fourier transform, we obtain matrix elements in the original momentum representation
[cf.\ Eq.\ \eqref{V0-munu}]:
\be
\label{Vmunu-ray}
  V_{\mu\nu} 
  = 
  \alpha\omega_0
  \left[
    \frac{h_{\mu\nu}}{\pi(\mu+\nu)} 
  - \frac{i}{2} \delta_{\mu+\nu} \sign\mu
  \right] .
\ee

\section{TWO INTERSECTING LINES}

\subsection{Spectrum for two perpendicular lines}
\label{cross}
 
The quasi-local Schr\"odinger equation for two perpendicular lines passing through the center of the vortex is given by Eq.\ \eqref{Schr-eq-2} with $\vp=\pi/2$. Similar to the single-line treatment in Sec. \ref{shred_eq_text}, the Schr\"odinger equation can be brought to a local form by arranging $\psi(\pm x)$ and $\psi(\pm(x-\pi))$ into the vector
\be
\Psi(x) = \begin{pmatrix}
\psi(x)\\
\psi(-x)\\
\psi(x - \pi)\\
\psi(-x + \pi)
\end{pmatrix}
.
\ee
It is sufficient to consider the evolution of $\Psi(x)$ at the interval $x \in [0,\pi/2]$, since its various components then span the whole circle $[0,2\pi]$. The $2\pi$ antiperiodicity of the wavefunction imposes the following constraints on $\Psi$ at the beginning and at the end of the interval $[0,\pi/2]$:
\be \label{structure}
\Psi(0) = \begin{pmatrix}
a\\
a\\
-b\\
b\\
\end{pmatrix},
\qquad
\Psi(\pi/2) = \begin{pmatrix}
c\\
d\\
d\\
c\\
\end{pmatrix}
.
\ee
Here $\psi(0) = a$, $\psi(\pi) = b$, $\psi(\pi/2) = c$ and $\psi(-\pi/2) = d$.

The evolution of $\Psi$ can be written as
\be \label{matrix_eq_90}
\partial_x \Psi(x) = M_E \Psi(x)
\ee
with
\be
M_E = \begin{pmatrix}
iE/\omega_0 & \alpha_1 & 0 & -\alpha_2 \\
\alpha_1 & -iE/\omega_0 & -\alpha_2 & 0 \\
0 & -\alpha_2 & iE/\omega_0 & -\alpha_1 \\
-\alpha_2 & 0 & -\alpha_1 & -iE/\omega_0
.
\end{pmatrix}
\ee
The vectors at the edges of the interval are related by the transfer matrix:
\be \label{evolution}
\Psi(\pi/2) = S_E(\pi/2) \Psi(0) ,
\ee
which becomes just a trivial matrix exponent since $M_E$ is $x$ independent: $S_E(\pi/2) = \exp(M_E\pi/2)$. After some algebra, we obtain
\be
S_E(\pi/2) = \cos(k\pi/2) + \frac{\sin(k\pi/2)}{\kappa} M_E ,
\ee
where $k = \sqrt{(E/\omega_0)^2 - \alpha_1^2-\alpha_2^2}$.
Processing now the constraints \eqref{structure}, we arrive at the following equation for the allowed momenta $k$:
\be
\label{lambda-eq}
  \frac{
    (k^2-\alpha_1\alpha_2)\cos\pi k
  + (\alpha_1+\alpha_2) k \sin\pi k
  + \alpha_1\alpha_2}{k^2}
  =
  0 .
\ee
This equation generalizes Eq.\ \eqref{eq-spectrum} to the two-line case and reduces to the latter if either $\alpha_1$ or $\alpha_2$ goes to zero. For given $\alpha_1$ and $\alpha_2$, transcendental Eq.\ \eqref{lambda-eq} defines a discrete set of momenta $k_n(\alpha_1,\alpha_2)>0$, which we label starting with $n=0$, as in Sec.\ \ref{zero_distance}.
The spectrum is then given by
\be
  E_n = \omega_0\sqrt{
\alpha_1^2+\alpha_2^2+k_n^2(\alpha_1,\alpha_2)} ,
\ee
where $k_n$ are defined by Eq.\ \eqref{lambda-eq}.

In the limit of strong defects ($\alpha_1^2+\alpha_2^2\gg1$), $k_0\approx2$ and the minigap takes the form
\be 
  E_g = \omega_0\sqrt{\alpha_1^2+\alpha_2^2} .
\ee

\subsection{Minigap asymptotics for $\vp=\pi/n$}
\label{AA:minigap-asympt}
 
A common feature of the one-line case considered in Sec.\ \ref{zero_distance} and the two-perpendicular-line case analyzed in Appendix \ref{cross} is that finding the asymptotic behavior of the minigap is much easier than determination of the whole spectrum. While the latter requires calculating discrete momenta by solving a transcendental spectral equation, those are not needed to compute the asymptotics. This observation immediately leads to the following criterion for the minigap determination: It is the first positive solution of
\be
\label{detM=0}
  \det M_E = 0 ,
\ee
where the matrix $M_E$ governs chiral evolution of $\Psi$ [see Eqs.\ \eqref{M-eq} and \eqref{matrix_eq_90}]. Equation \eqref{detM=0} also holds for any rational $\vp/\pi$ when the vector $\Psi$ is finite.

In the case $\vp=\pi/n$, the vector $\Psi$ has $2n$ components. We arrange them according to Eq.\ \eqref{mirror} and obtain the matrix $M_E$. In general, $M_E$ is a symmetric matrix with the following properties: (i) its main diagonal contains $\pm iE/\omega_0$ in alternating order, (ii) its 1-diagonal contains $\alpha_1$ and $\alpha_2$ in alternating order, with the elements ($2,3$) to ($n+1,n+2$) having an additional minus sign, (iii) the element ($1,2n$) equals $-\alpha_2$, (iv) other elements not related by the symmetry are zero.
The structure is illustrated by the $n=4$ example (here $\eps = E/\omega_0$):
\be
\nonumber
M_E
=
\begin{pmatrix}
i\eps & \alpha_1 & 0 & 0 & 0 & 0 & 0 &
   -\alpha_2 \\
 \alpha_1 & -i \eps & -\alpha_2 & 0 & 0
   & 0 & 0 & 0 \\
 0 & -\alpha_2 & i \eps & -\alpha_1 & 0
   & 0 & 0 & 0 \\
 0 & 0 & -\alpha_1 & -i \eps & -\alpha_2 & 0 & 0 & 0 \\
 0 & 0 & 0 & -\alpha_2 & i \eps & -\alpha_1 & 0 & 0 \\
 0 & 0 & 0 & 0 & -\alpha_1 & -i \eps &
   \alpha_2 & 0 \\
 0 & 0 & 0 & 0 & 0 & \alpha_2 & i \eps &
   \alpha_1 \\
 -\alpha_2 & 0 & 0 & 0 & 0 & 0 &
   \alpha_1 & -i \eps \\
  \end{pmatrix}
  .
\ee

Using determinant properties, the minigap equation \eqref{detM=0} can be equivalently written as
\be
\label{detN=0}
  \det (\eps \openone + i R) = 0 ,
\ee
where $\openone$ is a unit matrix and
\be
\nonumber
R
=
\begin{pmatrix}
 0 & \alpha_1 & 0 & 0 & 0 & \cdots & 0 & \alpha_2 \\
 - \alpha_1 & 0 & \alpha_2 & 0 & 0 &  & 0 & 0 \\
 0 & - \alpha_2 & 0 & \alpha_1 & 0 &  & 0 & 0 \\
 0 & 0 & - \alpha_1 & 0 & \alpha_2 &  & 0 & 0 \\
 0 & 0 & 0 & - \alpha_2 & 0 &  & 0 & 0 \\
 \vdots &  &  &  &  & \ddots &  & \vdots \\
 0 & 0 & 0 & 0 & 0 &  & 0 & \alpha_1 \\
 -\alpha_2 & 0 & 0 & 0 & 0 & \cdots & -\alpha_1 & 0
\end{pmatrix}
  .
\ee
This matrix is diagonalized in the momentum representation by a two-site modulated plane wave 
$u_a = w^{(-1)^a}e^{iq_sa}$ with $a=1,\dots,2n$. Solving for the modulation depth $w$, we obtain the spectrum of the matrix $iR$:
\be
  \lambda^2(q_s) = \alpha_1^2+\alpha_2^2-2\alpha_1\alpha_2\cos(2q_s) .
\ee
Momentum quantization is influenced by a ``wrong sign'' of the top-right matrix element of $R$ that results in $q_s=(s-1/2)(\pi/n)$ with $s=1,\dots,2n$. Thus, Eq.\ \eqref{detN=0} yields
\be
  \prod_{s=1}^n [E^2-\lambda^2(q_s)] = 0 .
\ee
The minimal positive solution of this equation is evidently $E=\lambda(q_1)$, leading to the minigap asymptotics~\eqref{Eg-asympt}.

The same analysis can be repeated for angles $\vp=(m/n)\pi$ with $m\neq1$. Position of plus and minus signs in front of $\alpha_1$ and $\alpha_2$ in the matrix $M_E$ will be different, but the matrix $R$ will be exactly the same.

\section{EFFECT OF $\Delta(\br)$ DISTORSION}\label{A:self_consistency}

In this Appendix, we estimate the effect of $\Delta(\br)$ modification on the quasiparticle spectrum. We perform the analysis in the simplest case of a sufficiently weak defect, $1\ll\alpha\ll\sqrt{k_F\xi}$, when the subgap states localized along the defect discussed in Sec.\ \ref{S:NN} do not appear and simple quantum mechanics developed in Sec.\ \ref{S:one-line} applies.
The zero-temperature limit is assumed.

\subsection{Perturbative correction to $\Delta(\br)$}

According to the self-consistency equation~\cite{deGen}, the order parameter is given by the sum over quasiparticle states:
\begin{equation}
\label{SCE}
\Delta(\mathbf{r}) 
= 
g\sum_{n} u_{n}(\br)v_n^{*}(\br) \tanh(E_n/2T) ,
\end{equation}
where $g$ is the BCS coupling constant, $T$ is temperature, and summation goes over positive energies, $E_n>0$. In Eq.\ \eqref{SCE}, $u$ and $v$ are the particle and hole components of the wave function.

Following the approach discussed in Sec.\ \ref{S:self_consistency}, we are going to determine the first approximation to the order parameter, $\Delta_1(\br)$, taking the wave functions in the presence of the linear defect but calculated with the clean $\Delta_0(\br)$.
Since the states of the chiral branch exhibit strong modification by the defect, we expect that they give the leading contribution to $\delta\Delta(\br)=\Delta_1(\br)-\Delta_0(\br)$.
Hence, we will replace the latter by
$
  \delta\Delta(\br)
  =
  \Delta_1^\text{ch}(\br)-\Delta_0^\text{ch}(\br) 
$,
where the right-hand side contains only the contribution of the chiral branch to the self-consistency equation \eqref{SCE}.
Replacing then $u$ and $v$ by the components of the real-space wave function \eqref{psi-real-space}, and we obtain at zero temperature:
\begin{multline}
\label{Delta-eq-1}
\Delta_1^\text{ch}(\br) 
= 
gA^2e^{-2K(r)}
\sum_{n}\sum_{\mu,\nu}
\corr{\mu|n}
\corr{n|\nu}
\\{}
\times J_{\mu - 1/2}(k_Fr)J_{\nu + 1/2}(k_Fr)
e^{i(\mu - \nu - 1)\vp},
\end{multline}
where $n$ labels the states in the presence of the defect and the overlaps $\corr{\mu|n}$ are given by Eq.\ \eqref{overlap}.

In order to obtain the correction to the order parameter, $\delta\Delta(\br)$, one should subtract the chiral-branch contribution in the clean ($\alpha=0$) case, leading to
\be
\label{deltaDelta}
  \delta\Delta(\br)
  =
  \Delta_1^\text{ch}(\br)
- \Delta_1^\text{ch}(\br) \big|_{\alpha=0} .
\ee

It is convenient to present Eq.\ \eqref{Delta-eq-1} as a sum over angular harmonics:
\be \label{delta_m_def}
\Delta_1^\text{ch}(\br) = gA^2 e^{-2K(r)}e^{-i\vp}\sum_{m = -\infty}^{\infty}
\gamma_m(r)e^{im\vp}.
\ee
Here, the $m$th harmonic of the order parameter $\gamma_m(r)$ can be represented as a sum of contributions from the overlaps with the $n$th state: $\gamma_m(r) = \sum_n \gamma_{mn}(r)$, where
\be \label{Delta_mn}
\gamma_{mn} = \sum_{\mu} \corr{\mu|n}\corr{n|\mu - m}J_{\mu - 1/2}(k_F r)J_{\mu - m + 1/2}(k_F r) .
\ee

In the clean case, only the zero harmonic is present, and the summation over $\mu$ and $n$ can be easily carried out:
\be
\label{gamma-0-clean}
\gamma_{0}(r)\big|_{\alpha=0}
= 
\frac{k_F r}{2}
\left[J_0^2(k_Fr) + J_1^2(k_Fr) \right] .
\ee

\begin{figure}
\centerline{\includegraphics[width=0.95\columnwidth]{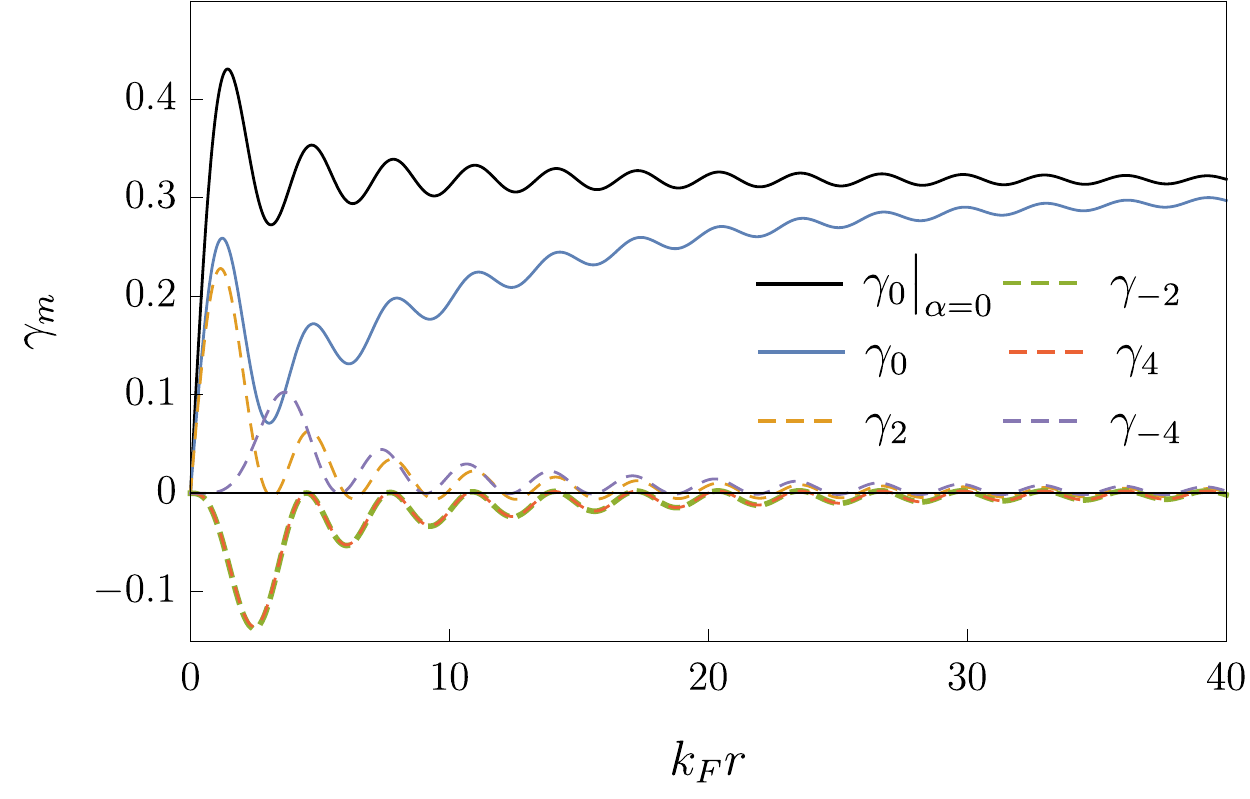}}
\caption{Angular harmonics of the chiral-branch contribution to the order parameter $\Delta_1^\text{ch}$ given by Eq. \eqref{delta_m_def} as a function of the distance from the vortex center. The defect strength $\alpha = 10$. The black curve represents $\gamma_0(r)$ in the clean case, see Eq.\ \eqref{gamma-0-clean}.}
\label{delta_harmonics}
\end{figure}

In the presence of the defect, nonzero harmonics appear. Due to the $\pi$-shift symmetry of the wave functions [Eq.\ \eqref{psi:x+pi}] odd harmonics vanish: $\gamma_{2k + 1}(r) = 0$. The profiles of several lowest harmonics are shown in Fig.\ \ref{delta_harmonics}.

\subsection{Back action on the spectrum}

Here we calculate the shift of the minigap due to the modification of the order parameter $\delta\Delta(\br)$ given by Eq.\ \eqref{deltaDelta}.
It can be obtained by treating the emerging correction to the BdG Hamiltonian \eqref{H-BdG} 
by the first-order perturbation theory:
\begin{equation}
\label{dE0-1}
  \Delta E_g
  = 
  \int d\br \,
  \Psi_{0}^\dagger(\br)
  \begin{pmatrix}
    0  & \delta \Delta(\br) \\
    \delta \Delta^{*}(\br) & 0\\
  \end{pmatrix} \Psi_{0}(\br) ,
\end{equation}
where $\Psi_0(\br)$ is the wave function with the lowest positive energy $E_g(\alpha)$ [see Eqs.\ \eqref{Eg-gen} and \eqref{psi-real-space}].

The shift of the minigap can be decomposed into the contributions due to the distortion of the zero-harmonic profile of the order parameter and due to its axial symmetry distortion by higher harmonics:
\be
\Delta E_g = \Delta E_g^{0} + \Delta E_g^{\neq 0}.
\ee
The expressions for these contributions can be written in the terms of the order parameter harmonics introduced in Eq.\ \eqref{Delta_mn} as
\begin{multline}
\label{energy_shift_zero}
\Delta E_g^{0} = \frac{4\pi gA^4}{k_F^2}\int_0^{\infty}dz\,z \, e^{-4K(z/k_F)} \gamma_{00}(z)
\\{}
\times \Bigl[ \gamma_0(z) - \gamma_0(z)\big|_{\alpha=0} \Bigr].
\end{multline}
and
\be\label{energy_shift_nonzero}
\Delta E_g^{\neq 0} = \frac{4\pi gA^4}{k_F^2}\int_0^{\infty}dz\,z\,e^{-4K(z/k_F)} 
\sum_{m\neq0}
\gamma_{m0}(z)\gamma_m(z).
\ee

The integrals above can be calculated numerically for different values of the parameter $\alpha$. It turns out that the $m=0$ contribution \eqref{energy_shift_zero} converges at small distances, therefore, the factor $e^{-4K(r)}$ can be omitted and $\Delta E_g^{0}$ appears to be independent of $k_F\xi$:
\be
  \Delta E_g^{0} = \frac{4\pi gA^4}{k_F^2} c_0(\alpha) ,
\ee 
where the coefficient $c_0(\alpha)$ should be determined numerically.
On the other hand, Eq.\ \eqref{energy_shift_nonzero} has a logarithmic divergency at large distances, and the factor $e^{-4K(r)}$ provides an infrared cutoff at $z \sim k_F\xi$:
\be
  \Delta E_g^{\neq 0} = \frac{4\pi gA^4}{k_F^2} c_1(\alpha) \ln k_F\xi ,
\ee
where the coefficient $c_1(\alpha)$ should be determined numerically.

\begin{figure}
\centerline{\includegraphics[width=0.95\columnwidth]{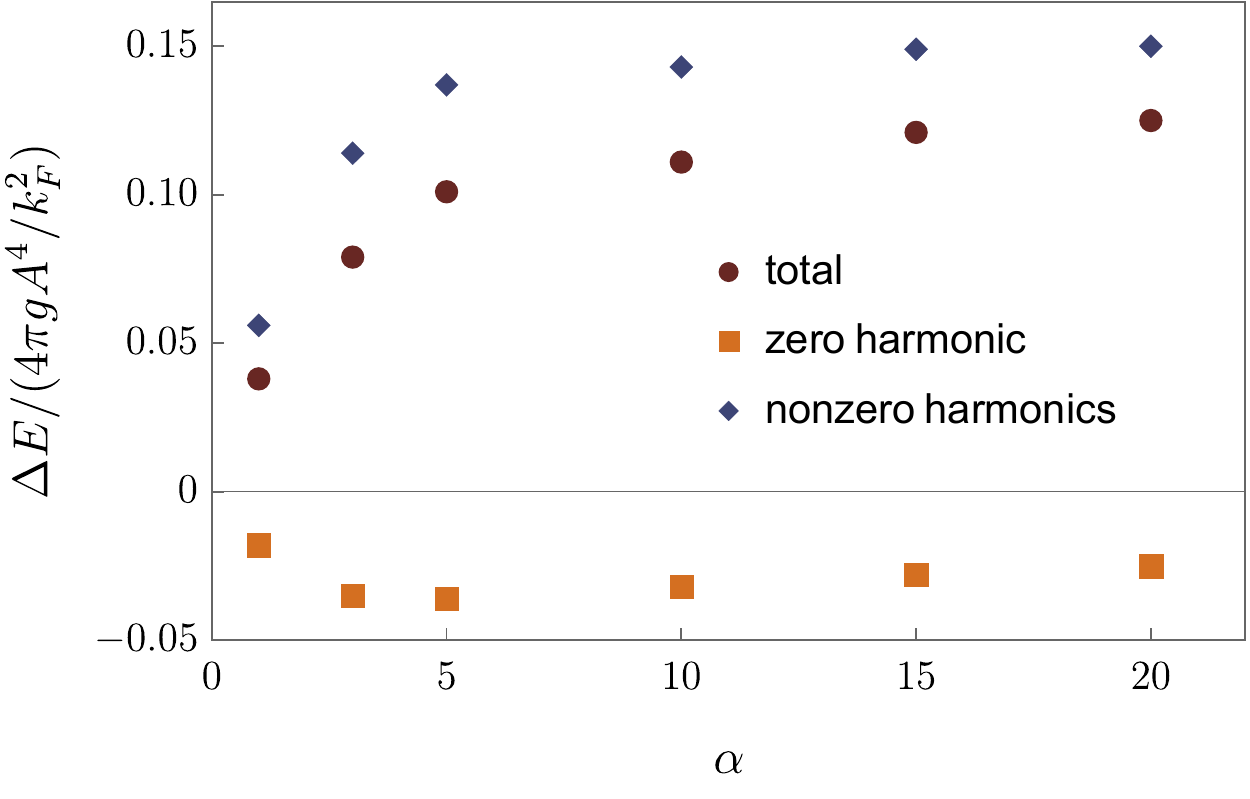}}
\caption{The shift of the minigap $\Delta E_g$ for several values of the defect strength $\alpha$ obtained numerically for $k_F\xi = 500$. The total value of the shift is depicted by brown circles. Orange squares and blue diamonds represent the contributions from the zero ($\Delta E_g^0$) and nonzero ($\Delta E_g^{\neq0}$)  harmonics of the order parameter, correspondingly.}
\label{energy_shifts}
\end{figure} 

Figure \ref{energy_shifts} represents the shift of the minigap $\Delta E_g$ for several values of the defect strength $\alpha$, as well as its contributions from zero ($\Delta E_g^0$) and nonzero ($\Delta E_g^{\neq0}$) harmonics. The calculations were performed at $k_F\xi=500$. We see that the growth of $\Delta E_g$ nearly saturates for $\alpha>10$ at the value of $c_0+c_1\ln k_F\xi\approx 0.1$.

\end{document}